\RequirePackage{fix-cm}
\documentclass[natbib,smallextended]{svjour3}       
\smartqed  
\usepackage{graphicx}
\usepackage{multicol}
\usepackage{float}

\usepackage{amsmath}
\usepackage{amssymb}

\usepackage{color,soul}

\usepackage{mathptmx}      
\usepackage{latexsym}
 \journalname{my journal}
\begin{document}

\title{Exoplanetary Atmospheres -- Chemistry, Formation Conditions, and Habitability}


\author{Nikku Madhusudhan         \and Marcelino Ag\'{u}ndez        \and
	     Julianne I. Moses         \and
              Yongyun Hu         \and
}


\institute{N. Madhusudhan \at
                Institute of Astronomy, Madingley Road \\
                 University of Cambridge, Cambridge CB3 0HA, UK \\
              \email{nmadhu@ast.cam.ac.uk}           
           \and
           M. Agundez Chico \at
           Instituto de Ciencia de Materiales de Madrid, \\ 
           CSIC, C/ Sor Juana InŽs de la Cruz 3, 28049 Cantoblanco, Spain \\
           \email{marcelino.agundez@icmm.csic.es} 
           \and
           J. I. Moses \at
           Space Science Institute, \\ 
           4750 Walnut Street, Suite 205, Boulder, CO 80301, USA \\ 
           \email{jmoses@SpaceScience.org} 
            \and 
            Y. Hu \at  
            Laboratory for Climate and Ocean-Atmosphere Sciences, \\ 
            Department of Atmospheric and Oceanic Sciences, School of Physics, \\ 
            Peking University, Beijing 100871, China \\
             \email{yyhu@pku.edu.cn}
}

\date{Received: July 29, 2015 / Accepted: March 12, 2016}

\maketitle


\begin{abstract}
Characterizing the atmospheres of extrasolar planets is the new frontier in exoplanetary science. The last two decades of exoplanet discoveries have revealed that exoplanets are very common and extremely diverse in their orbital and bulk properties. We now enter a new era as we begin to investigate the chemical diversity of exoplanets, their atmospheric and interior processes, and their formation conditions. Recent  developments in the field have led to unprecedented advancements in our understanding of atmospheric chemistry of exoplanets and the implications for their formation conditions. We review these developments in the present work. We review in detail the theory of atmospheric chemistry in all classes of exoplanets discovered to date, from highly irradiated gas giants, ice giants, and super-Earths, to directly imaged giant planets at large orbital separations. We then review the observational detections of chemical species in exoplanetary atmospheres of these various types using different methods, including transit spectroscopy, doppler spectroscopy, and direct imaging. In addition to chemical detections, we discuss the advances in determining chemical abundances in these atmospheres and how such abundances are being used to constrain exoplanetary formation conditions and migration mechanisms. Finally, we review recent theoretical work on the atmospheres of habitable exoplanets, followed by a discussion of future outlook of the field. 

\end{abstract}


\section{Introduction}
\label{sec:intro}
The study of extrasolar planets (or `exoplanetsÕ) is one of the most dynamic frontiers of modern astronomy. A few thousand exoplanets are now known. The numerous exoplanet discoveries via different methods have shown that exoplanetary systems are very common and are extremely diverse in their macroscopic properties. Recent statistics from observational surveys are suggesting that almost every star in the solar neighborhood hosts at least one planet around it, and that terrestrial-size exoplanets are amongst the most numerous \citep{Howard2012,Fressin2013}. The exoplanets detected to date span a diverse range in masses, radii, temperatures, and orbital parameters (orbital periods, separations, eccentricities, inclinations, etc.). The extreme diversity of these macroscopic properties of exoplanets has for the first time placed the solar system in cosmic context, thereby opening a plethora of new questions with far reaching implications. 

We are now entering a new era in exoplanetary science. Going beyond planet detections, the field is now moving towards detailed characterization of exoplanetary atmospheres through spectroscopic observations. Encoded within a spectrum of an exoplanet is information about the chemical composition and manifold physical processes in its atmosphere. State-of-the-art observations are now beginning to provide both the high sensitivity and long spectral baseline required to place detailed constraints on the various physicochemical properties of exoplanetary atmospheres. Such observations in recent years are already providing good constraints on the chemical compositions of exoplanetary atmospheres. The most observed exoplanets to date are hot giant planets whose large scale-heights and high temperatures ($\sim$800-3000 K) make them particularly conducive to atmospheric observations. Several prominent molecules of carbon and oxygen and several atomic species (e.g. Na, K) are expected to be abundant and observable in these atmospheres, making giant exoplanets rich laboratories for understanding atmospheric chemistry. The atmospheric chemical compositions are in turn beginning to provide the first insights into the possible formation conditions and migration mechanisms of exoplanets. 

Atmospheric observations have been reported for a variety of exoplanets detected via transits, direct imaging, as well as the radial velocity method. Spitzer observations have been obtained for about 20 transiting exoplanets in at least four photometric bands (3.6, 4.5, 5.8, 8 $\mu$m) and about 50 in two (3.6 and 4.5 $\mu$m), primarily for close-in hot Jupiters \citep{madhusudhan2014a}. Photometric observations have also been obtained in the near infrared using large ground-based telescopes. Recently, high S/N near-infrared transit spectroscopy has become possible thanks to the HST Wide Field Camera 3 (WFC3) leading to the first high-confidence detections of H$_2$O in several transiting exoplanets (e.g. Deming et al. 2013). Multi-wavelength datasets are also providing the long spectral baseline and high precision required to derive joint constraints on the chemical compositions and temperature profiles of exoplanetary atmospheres. Over 400 HST orbits have been recently allocated for high S/N spectra of over a dozen transiting exoplanets using the HST instruments in the visible and near-infrared. High resolution spectroscopic observations have also been reported for several directly imaged planets in the near-infrared, and dedicated surveys are expected to pursue the same for dozens more. Finally, it has also now become possible to detect molecules in the atmospheres of close-in planets using very high resolution ($R = 10^5$) doppler spectroscopy \citep{snellen2010}. These observational advancements have led to the detection of chemical species in a sizable ensemble of planets and have motivated rapid progress in our theoretical understanding of atmospheric chemistry in exoplanets. 

In the present work, we review the latest developments in our understanding of atmospheric chemistry of exoplanets. We begin with a brief introduction to observational methods for exoplanet detection and atmospheric characterization in section~\ref{sec:obsmethods}. We then discuss the theoretical developments in our understanding of atmospheric chemistry in exoplanets (in section~\ref{sec:chemistry}), with particular emphasis on large gas giant planets (in section~\ref{sec:chemistry_gas_giants}) as well as ice giants and super-Earths (in section~\ref{sec:chemistry_nep}). We then review, in section ~\ref{sec:inferences}, observational inferences of chemical species in exoplanetary atmospheres of various class using different observational methods. In section~\ref{sec:formation}, we review developments towards a theoretical framework for using atmospheric chemical abundances to constrain exoplanetary formation conditions and migration pathways. In section~\ref{sec:habitability}, we discuss theoretical developments in our expectations for the habitability of terrestrial exoplanets. We conclude with a discussion of the future outlook for the field. 

\section{Observational Methods}
\label{sec:obsmethods}
The thousands of exoplanets known to date have been detected via a variety of observational methods \citep[see][for a review]{fischer2014}. By far the majority of the planets have been detected using the transit method and the radial velocity method, followed by some detections using various other methods such as pulsar timing, gravitational microlensing, and direct imaging. On the other hand, observations of exoplanetary atmospheres present much more stringent requirements on the instrumental capabilities than planet detection. A combination of significantly higher sensitivity and higher spectral resolution are required for reliable atmospheric detections. Amongst the various planet finding methods, those amenable for atmospheric characterization are the transit method, direct imaging, and radial velocity method, with the transit method being the most successful to date. A detailed review of all these observational methods for atmospheric characterization can be found in \citet{madhusudhan2014a}. Here we briefly review the key attributes of each method. 

One of the most successful methods for observing exoplanetary atmospheres has been the transit method. When the planet transits in front of the host star, part of the star light traverses through the day-night terminator region of the planetary atmosphere before reaching the observer. The resulting  `transmission spectrum', obtained by subtracting the in-transit spectrum from the out-of-transit spectrum, contains absorption features imprinted on the starlight by chemical species in the planetary atmosphere. Therefore, transmission spectra probe the chemical composition and temperature structure at the day-night terminator of the planet. On the other hand, when the planet is at full phase before being occulted by the star (i.e. 'secondary eclipse') the thermal emission and reflection spectrum from the planet is observed along with the stellar spectrum which can be subtracted out later; during secondary eclipse only the stellar spectrum is observed. Thus, secondary eclipse spectra probe composition and thermal structure of the dayside atmosphere of the planet. Spectra for exoplanets have been obtained in both transmission and at secondary eclipse, and in a wide range of wavelengths. While transmission spectra have been obtained from UV to mid-infrared wavelengths, thermal emission spectra have been obtained predominantly in the near to mid infrared. Besides the transit method, atmospheric observations have also been obtained from direct imaging of planets as well as by high-resolution doppler spectroscopy, which are both discussed briefly in their respective sections below. 

\section{Factors Influencing Atmospheric Chemistry of Exoplanets}
\label{sec:chemistry}

The advent of observational techniques able to characterize atmospheres of extrasolar planets has been a major breakthrough in astronomy. The discovery of a great diversity of exoplanets, many of them with no analogue in the solar system, suggests that there is an exoplanet zoo out there with a continuum of possible sizes (from Jupiter-sized planets to sub-Earth bodies), atmospheric temperatures (from thousands of degrees Kelvin down to a few tens), and elemental compositions (from H/He-dominated atmospheres with a solar composition to atmospheres dominated by heavy molecules such as H$_2$O, CO$_2$, or N$_2$). 
This diversity poses a great challenge to theorists who seek an understanding of the variety of existing planetary climates.

A theoretical approach to the chemistry of exoplanet atmospheres must start by posing a question of the following type.  How can we infer the chemical composition of a planetary atmosphere from a reduced set of parameters related to the planet$?$ Of course it is first necessary to identify such parameters and to understand how each of them affect the
atmospheric chemistry. Some of these parameters can be obtained in a straightforward way from observations, although there are some others for which we can just have guesses, albeit in some cases with formidable uncertainties. Let's have a look to each of these parameters.

\emph{Gravity.--} This parameter follows directly from the mass and radius of the planet, which in turn can be derived from radial velocity and transit techniques, respectively. Whether or not a planet can retain a substantial amount of atmosphere is largely related to its gravity and the X-ray and EUV flux of the host star (see e.g. chapter on protoatmospheres in present issue, Massol et al. 2016). Moreover, depending on the mass and radius of the planet, light elements such as hydrogen and helium may or not escape from the planet, with strong implications for the elemental composition of the atmosphere. The gravity is also important in that, together with the mean mass of particles and temperature, it sets the scale height of the atmosphere, i.e., how compact or extended the atmosphere is.

\emph{Elemental composition.--} The relative abundances of the different elements is one of the most important aspects that determine the atmospheric composition, although unfortunately these are rather difficult to infer from observations. Based on the study of solar system planets and on existing theories of planet formation \citep{1993ARA&A..31..129L,2002ARA&A..40..103H,2004ARA&A..42..549G,2005A&A...434..343A,2006RPPh...69..119P,2008ARA&A..46...21B,2010AREPS..38..493C,2012AREPS..40..251M},
it is expected that giant planets will from by core accretion, becoming massive enough at an early stage of planet
formation to efficiently capture the nebular gas, and thus will retain a thick H/He-dominated atmosphere (see e.g. chapter on protoatmospheres in pressent issue, Massol et al. 2016).  Terrestrial planets, on the other hand, will not become massive enough to efficiently accrete or retain H$_2$/He from the nebular gas, and their atmospheres, whether thick or thin, will depend on their mass and evolutionary history (e.g., orbital evolution, stellar/disk evolution, impact history, interior outgassing, atmospheric escape, climate evolution, surface-atmosphere interactions, magnetospheric interactions). Terrestrial-planet atmospheres are expected to be typically dominated by secondary products such as H$_2$O, CO$_2$, CO, N$_2$, Ne, Ar, Kr, SO$_2$, SiO$_2$ that were outgassed from the solid planetesimals that formed the planets or that were supplied by later impacts by solid bodies (see e.g. chapter on planet formation in present issue, Bai et al. 2016). In between giant and 
terrestrial planets there is a regime of super-Earths/sub-Neptunes whose elemental composition is largely unknown 
\citep[e.g.,][]{2013AREPS..41..469H,fortney13,lopez14,marcy14,lee14,rogers15}. By comparing the outcomes of chemical 
models and observations, it is in principle possible to put constraints on the elemental composition of the planetary 
atmosphere \citep[e.g.,][]{madhusudhan2014a}, which in turn can provide clues to the formation and evolution of the planet itself.

\emph{Insolation.--} The amount of energy received at the top of the atmosphere per unit time and unit area depends essentially
on the luminosity of the host star and the orbital distance, two properties that can be relatively well constrained by
observations. Of course, the way the stellar luminosity is divided across the electromagnetic spectrum, i.e., the spectral type
of the star, has a great importance in affecting atmospheric characteristics. Incoming visible and infrared photons take 
care of the heating of the atmosphere, especially at pressures greater than $\sim$1 microbar
\citep[e.g.,][]{seager98,sudarsky2003,barman05,2005A&A...436..719I,fortney05,burrows06}, 
while ultraviolet photons heat the high-altitude thermosphere and lead to ionization and dissociation of atmospheric constituents and induce photochemistry at a variety of altitudes \citep[e.g.,][]{2003ApJ...596L.247L,2004Icar..170..167Y,2007P&SS...55.1426G,2009arXiv0911.0728Z,2010ApJ...717..496L,2011ApJ...737...15M,2012A&A...546A..43V,2012A&A...548A..73A,hu12bench,miller-ricci12,2013Icar..226.1678K,lavvas14alkali,shaikhislamov2014,rimmer2015}. 

\emph{Internal heating.--} The interiors of planets can provide an important source of heat for the atmosphere, depending on the planet's age and tidal interactions with the host star or other planets in the system. The internal heating can be estimated from theoretical models of the evolution of the planet interior and its reaction to tidal forces \citep{2005AREPS..33..493G,2010RPPh...73a6901B,2014ARA&A..52..171O}.  
However, uncertainties in some key magnitudes, such as the dissipation properties of the bulk material of the planet interior and the age of the star (and thus of the planetary system), keep the internal heating poorly constrained in most cases. 

A self-consistent model of a planetary atmosphere should in principle be able to describe the physical and chemical state of the
atmosphere from the parameters described above. The situation resembles that of stellar atmospheres 
\citep[e.g.,][]{kurucz79,hauschildt99}, where models need just three
parameters (gravity, metallicity, and effective temperature) --- external insolation being not important. Of course, in certain
types of planets, parameters or processes different from those listed above, such as, e.g., exchange of matter with the surface
in the case of terrestrial planets \citep[e.g.,][]{leconte14} or grain formation and resulting opacity effects
\citep[e.g.,][]{tsuji96,marley96,allard97}, may be also of great relevance in establishing the physical and chemical atmospheric properties.

In practice, models of planetary atmospheres lack a full self-consistency and tend to focus on some particular aspects, adopting approximations for some others. On the one hand, there are one-dimensional models that concentrate on solving the radiative transfer and thermal balance in the vertical direction. These models result in a pressure-temperature ($P$-$T$) profile for which the temperature gradient ($dT/dP$) is subadiabatic in the radiative zone of the atmosphere and adiabatic in the deeper convective regions \citep[e.g.,][]{1989Icar...80...23M,1997ApJ...491..856B,seager2005,burrows2008}. These radiative-convective equilibrium models usually make some assumptions about the chemical composition of the atmosphere, either adopting a particular one suitable for the modeled planet or assuming that chemical equilibrium holds in the atmosphere, and neglecting any fluid dynamics. For an assumed composition, in addition to the $P$-$T$ profiles these non-gray numerical models also calculate the line-by-line thermal emission and reflectance spectra of the planetary atmospheres. Recent studies have also reported analytic and semi-analytic $P$-$T$ profiles derived for irradiated planetary atmospheres in radiative equilibrium under the assumptions of gray/non-gray opacities \citep[][]{hansen2008,guillot2010,robinson2012,heng2012b,parmentier2014,parmentier2015}. 

Other type of models, the so-called general circulation models (GCMs), were originally developed to study the climate of the Earth 
\citep[e.g.,][]{1956QJRMS..82..123P,1982RvGSP..20..851M} and are now routinely used to study atmospheres of other solar system
planets \citep[e.g.,][]{1999JGR...10424155F} as well as exoplanets such as hot Jupiters \citep[e.g.,][]{2002A&A...385..166S,2009ApJ...699..564S}, hot
Neptunes \citep[e.g.,][]{2010ApJ...720..344L}, super-Earths \citep[e.g.,][]{2011ApJ...743L..36C} and terrestrial planets
\citep[e.g.,][]{2013cctp.book..277S}. These are three-dimensional models which solve the Navier-Stokes equations or a reduced set of ``primitive equations'' (see e.g. \citealt{showman10,heng15}) and end up with a three-dimensional view of the circulation and thermal structure of the atmosphere. Some of the limitations of GCMs have to do with the approximations used to deal with the radiative transfer \citep[e.g.,][]{2009ApJ...699..564S} and the chemical processes \citep{2006ApJ...649.1048C}. Another type of models are those focused on the atmospheric chemistry, in which we concentrate hereafter.

Chemical models of planetary atmospheres aim at describing how the atmospheric constituents are distributed, usually in the
vertical direction. To build such a model it is necessary to have information on some basic parameters, among them three of the four aforementioned: (1) \emph{gravity}, (2) \emph{elemental composition}, and (3) \emph{ultraviolet/X-ray irradiation}. It is also necessary to have information on a couple of additional parameters.

(4) \emph{Thermal structure.--} The spatial distribution of the atmospheric temperature, at least in the vertical direction, can be retrieved from radiative-convective models and GCMs (see above). In the case of these latter models it is possible to get a three-dimensional map of the temperature. The atmospheric temperature has an enormous influence on the chemical composition because depending on whether the atmosphere is hot or cold the major constituents and the type of condensates formed (if any) are completely different. 

(5) \emph{Strength of transport processes.--} An atmosphere is in essence a fluid and therefore a variety of processes such as
advection, diffusion, and turbulent motions can occur at different scales, having as consequence the transport and mixing of
material between different regions. These processes are an important source of disequilibrium because they can transport
molecules that were originally formed in a given location to other atmospheric regions. In particular, the vertical chemical
structure can be strongly affected by transport processes, the strength of which is usually parameterized through an eddy
diffusion coefficient ($K_{zz}$ if referred to the vertical direction). It is, however, difficult to obtain realistic estimates
of the coefficient $K_{zz}$. In the case of solar system planets such as Jupiter and Saturn, if observations are able to
determine the abundance of certain species at a given altitude, it is possible to put constraints on the eddy diffusion
coefficient \citep[e.g.,][]{atreya86}. In the case of extrasolar planets, information on the strength of mixing
processes must be obtained from theoretical models such as GCMs \citep[e.g.,][]{2013A&A...558A..91P}.

If one wants to build a chemical model of a planetary atmosphere, these five ingredients (\emph{gravity}, \emph{elemental composition}, \emph{ultraviolet/X-ray irradiation}, \emph{thermal structure}, and \emph{strength of transport processes}) should be in principle enough to provide a good description of the chemical structure of the atmosphere.\\

The characterization of exoplanet atmospheres through observations has been to date restricted to hot Jupiters, some hot
Neptunes, and a few super-Earths, most of them characterized by transit techniques, and to a few young and self-luminous gas
giant planets with an orbital distance sufficiently large to be accessible by direct imaging. It is therefore not surprising that
theoretical studies of exoplanet atmospheres carried out to date have been to a large extent biased toward these types of
extrasolar planets.  A general theoretical characterization of habitable-zone exoplanet atmospheres has also been 
a popular topic, due to the high intrinsic interest in the potential for life (and its detectability) outside of Earth.

\section{Theory of Atmospheric Chemistry of Gas Giant Exoplanets}
\label{sec:chemistry_gas_giants}

\begin{figure*}
\includegraphics[width=\textwidth]{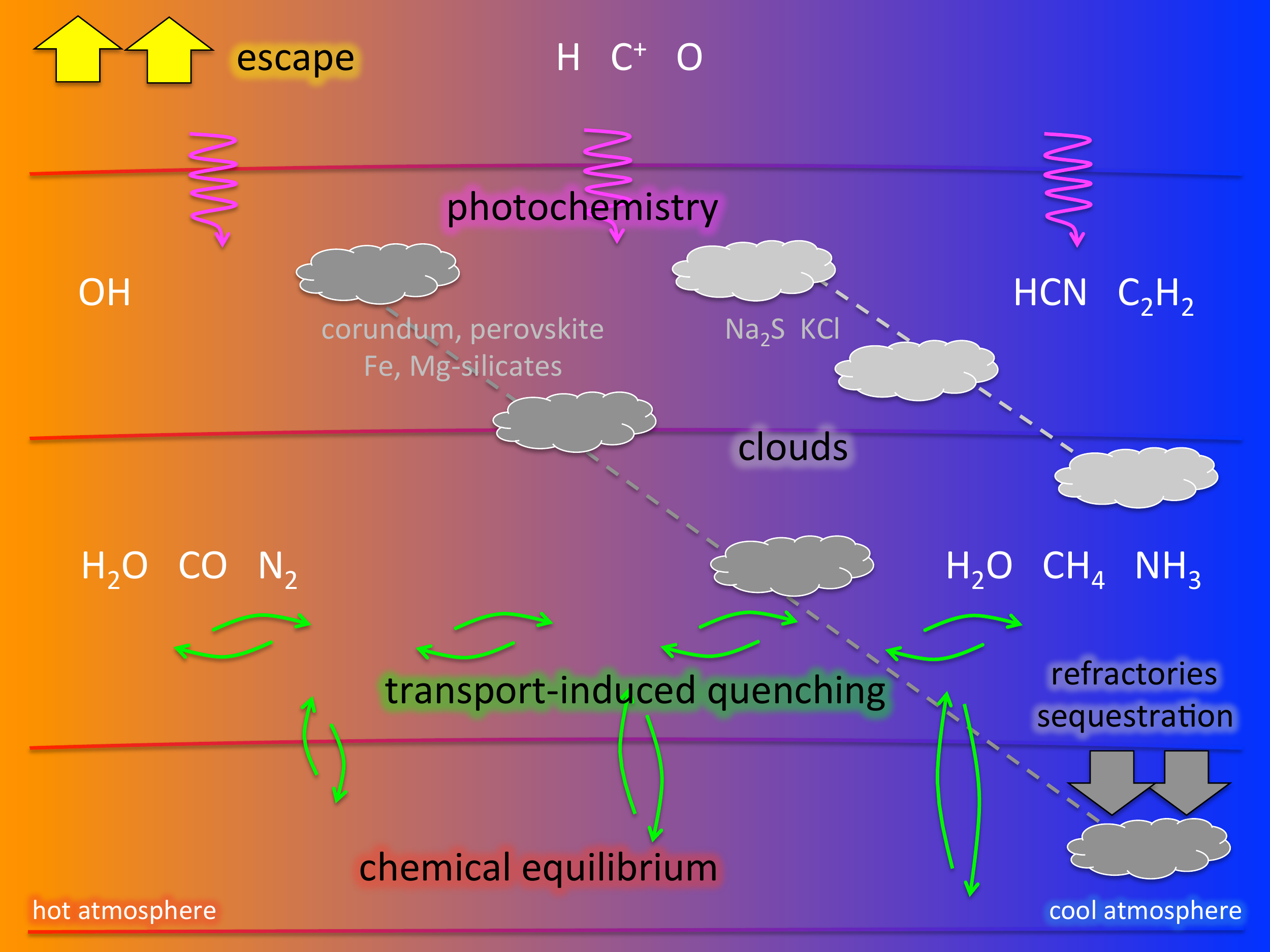}
\caption{Sketch of the structure and processes at work in H/He-dominated atmospheres. The chemical composition in deep layers is
controlled by chemical equilibrium, by transport-induced quenching in upper layers, and by photochemistry in still upper layers.
Formation of clouds is an issue of increasing importance as the atmosphere becomes cooler. Escape can also be an important
process in highly irradiated atmospheres (i.e., those with hot thermospheres), while sequestration of refractory elements becomes important in the cooler atmospheres.}
\label{fig:sketch}
\end{figure*}

Gas giant exoplanets are expected to have H/He-dominated atmospheres based on their formation mechanism and on the bulk densities inferred from observations \citep{lissauer07,2010RPPh...73a6901B}. Their chemical composition depends on
various parameters, perhaps the most important being the temperature. A sketch of the structure of such an atmosphere is depicted
in Fig.~\ref{fig:sketch}, where various important processes at work are indicated. The chemical composition is controlled by
chemical equilibrium in deep atmospheric layers, by transport-induced quenching in upper layers, and by photochemistry in still
upper layers. Cloud formation becomes increasingly important as the atmosphere is cooler. Atmospheric escape can be an important
issue for highly irradiated giant planets \citep[e.g.,][]{koskinen14esc}, while sequestration of refractory elements plays an 
important role in cool atmospheres \citep[e.g.,][]{lodders10}.

\subsection{Chemical Equilibrium}
\label{subsec:equilibrium}

Under the chemical equilibrium assumption, the chemical composition does only depend on temperature, pressure, and elemental
abundances, and can be calculated in a relatively straightforward way by minimizing the Gibbs free energy of the system. Chemical
equilibrium can accurately describe the composition of an atmosphere provided it is sufficiently hot and dense to ensure that
chemical reactions occur faster than any other process at work (dynamics, interaction with energetic radiation, etc.). This can
be the case for atmospheres of cool stars and brown dwarfs, where chemical equilibrium has been routinely applied
\citep{1973A&A....23..411T,1996ApJ...472L..37F,allard96,1999ApJ...512..843B,2002Icar..155..393L}, and of hot exoplanets with atmospheric temperatures in excess of $\sim$2000 K \citep{2011ApJ...737...15M,2012A&A...546A..43V}. In the case of cooler planets, it can still be a good starting point to have a general idea of which could be the main atmospheric constituents.

\begin{figure*}
\centering
\includegraphics[width=0.9\textwidth]{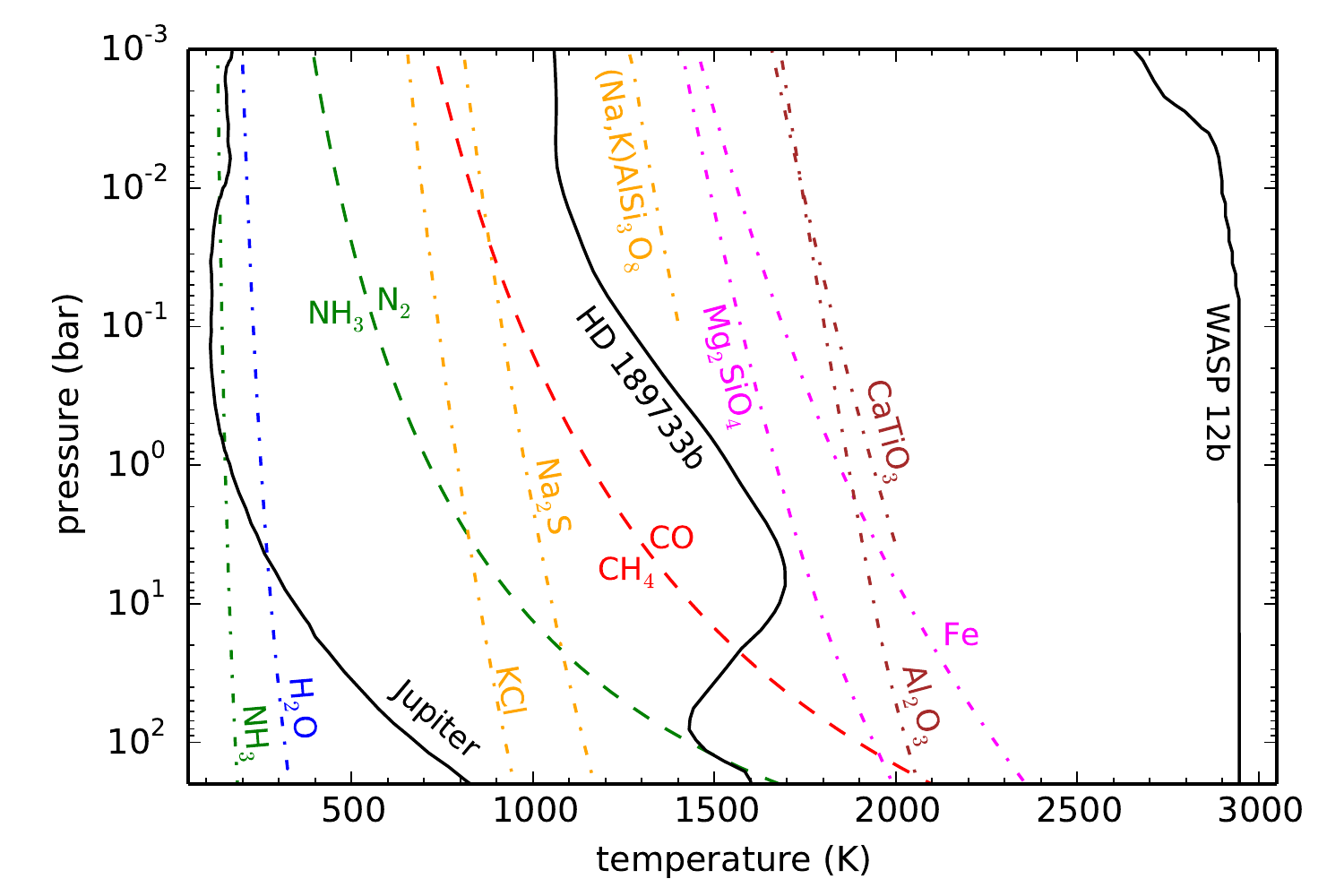}
\caption{Pressure-temperature profiles for three gas giant planets covering a broad range of atmospheric temperatures: Jupiter
\citep{1992AJ....103..967L}, the widely studied hot Jupiter HD~189733b \citep{2014A&A...564A..73A}, and the highly irradiated gas
giant WASP-12b \citep{stevenson2014a}. Dashed lines delimitate the regions where either CH$_4$ or CO (red curve)
 or NH$_3$ or N$_2$ (green curve) are the major carbon or nitrogen reservoirs, respectively. Dot-dashed lines indicate the condensation curves for water, ammonia, and various refractory species containing Ca, Ti, Al, Fe, Mg, Na, and K.}
\label{fig:pt_diagram}
\end{figure*}

The atmospheric temperatures of gas giant planets may span over a very broad range, from a few hundreds of kelvins up to
some thousands (see Fig.~\ref{fig:pt_diagram}). Gas giant planets are expected to have a nearly solar elemental composition,
although some elements heavier than helium could be enriched or depleted by a factor of a few relative to the solar composition,
as occurs in Jupiter \citep{mahaffy00,wong04}. In any case, gas giant atmospheres are expected to be dominated by hydrogen and
helium, with heavier elements like oxygen, carbon, and nitrogen being present at a lower level. Fig.~\ref{fig:pt_diagram} shows
various curves, calculated under chemical equilibrium for a solar elemental composition, which delimitate different regions of
interest from a chemical view point. The red dashed curve indicates where CO and CH$_4$ have the same abundance. To the right of
this curve (at high temperatures/low pressures) we have the region of stability of CO while to the left (at low temperatures/high
pressures) methane is the main carbon reservoir. Similarly, in the case of nitrogen chemical equilibrium indicates that to the
right of the green dashed curve N$_2$ is the main nitrogen reservoir while to the left it is ammonia that dominates. Thus, a
first lesson to be learnt from chemical equilibrium is that at high temperatures and/or low pressures CO and N$_2$ are the main
reservoirs of carbon and nitrogen, respectively, while at low temperatures and/or high pressures the hydrides CH$_4$ and NH$_3$
dominate. In the case of oxygen, water vapor remains a major reservoir over most of the temperature regime of interest. This
molecule locks either the excess of oxygen not locked into CO if the atmosphere resides in the stability region of CO (oxygen 
has a solar abundance about twice than carbon), or directly locks most of the available oxygen if the atmosphere resides in the 
stability region of CH$_4$. Only at very low temperatures (see blue dot-dashed line) can water vapor be depleted from the gas 
phase because of condensation, something that can occur also for ammonia, albeit at even lower temperatures (see green dot-dashed 
line). That is, according to chemical equilibrium, depending on the thermal profile, the atmosphere of a giant planet may be 
dominated by (apart from H$_2$ and He) a mixture of H$_2$O/CO/N$_2$, H$_2$O/CH$_4$/N$_2$, H$_2$O/CH$_4$/NH$_3$, or a modification 
of the latter in which first H$_2$O, then NH$_3$, and finally CH$_4$ are progressively removed from the gas phase as the
atmosphere gets cooler \citep[e.g.,][]{1997ApJ...491..856B,sudarsky00,2002Icar..155..393L}.

\subsection{Disequilibrium chemistry driven by transport processes}
\label{subsec:quenching}

The atmospheric composition of gas giant exoplanets is governed by chemical equilibrium only in the hottest regions, usually in
the deepest, densest regions (or the bulk of the atmosphere for very hot, strongly irradiated planets).  In cooler upper regions, when chemical reactions become slower than dynamic processes (advection, diffusion, or turbulence), the bulk composition can be significantly driven out of equilibrium. The transport of material between different regions has as a consequence a redistribution (homogenization) of heat and chemical composition, with a net flow from hot to cooler regions.

The vertical distribution of the atmospheric constituents can be strongly affected by the so-called chemical quenching, in which as material moves upwards chemical reactions become slower, and at a certain level the abundances of the different species are progressively quenched. This phenomenon was first described by \cite{1977Sci...198.1031P} to explain the presence of CO in the troposphere of Jupiter as a result of upward mixing from deeper and hotter levels, where CO is thermochemically stable. In gas giant exoplanets, chemical quenching regulates to a large extent the composition of the layers probed by observations, especially if their pressure-temperature profiles cross the boundaries of stability of the pairs CH$_4$/CO or NH$_3$/N$_2$ (see Fig.~\ref{fig:pt_diagram}). A quantitative evaluation of chemical quenching (i.e., the quench level and resulting quenched abundance) requires a good knowledge of both the chemical kinetics of the interconversion scheme of interest and the vertical eddy mixing coefficient.

The main issue when dealing with quenching and chemical conversion schemes is the identification of the elementary reactions 
involved and the ``limiting step'', whose rate controls the kinetics of the overall conversion. The interconversion CO 
$\rightleftharpoons$ CH$_4$ is of great importance in cool to moderately warm gas giant planets because exoplanet 
temperature-pressure profiles often cross the stability boundary between CO and CH$_4$ (red dashed curve in 
Fig.~\ref{fig:pt_diagram}).  Understanding how the conversion between CO $\rightleftharpoons$ CH$_4$ proceeds kinetically and 
which reaction is the rate-limiting step is thus of primordial importance (see \citealt{2011ApJ...738...72V} for more details). 
In the same vein, the interconversion N$_2$ $\rightleftharpoons$ NH$_3$ is also of great relevance for the atmospheres of gas 
giant planets, N$_2$ and NH$_3$ being the main reservoirs of nitrogen at high and low temperatures (or low and high pressures), 
respectively. Although there 
are still important uncertainties in the reaction schemes and the rate-limiting step, the conversion N$_2$ $\rightarrow$ NH$_3$ 
is likely intrinsically slower than the conversion CO $\rightarrow$ CH$_4$, resulting in a deeper quench level for the nitrogen 
species (see \citealt{2014RSPTA.37230073M} for more details). The identification of interconversion reaction schemes is interesting 
in that it permits implementation of simple prescriptions to deal with chemical quenching in high-demanding computational models of 
atmospheres such as GCMs \citep{2006ApJ...649.1048C}.

For 1D exoplanetary atmosphere models, the eddy diffusion coefficient must be quantified from 3D theoretical models like GCMs, or constrained from the observations themselves. Although mixing in planetary atmospheres can occur through large-scale advection, atmospheric waves, eddies of a variety of scales, and other transport processes that are not diffusive in a rigorous sense, vertical mixing can typically be well represented by a diffusion equation \citep[e.g.][]{lindzen81,strobel81,brasseur99}. A common practice has been to 
estimate the vertical diffusion coefficient $K_{zz}$ as the root mean square of the vertical velocity (as extracted from a GCM) 
times the vertical scale height \citep{2010ApJ...717..496L,2011ApJ...737...15M}. A better approach when 3D circulation models 
are available, as outlined by \citet{2013A&A...558A..91P}, is (1) to follow the behaviour of passive tracers in a GCM 
simulation and fit the resulting planet-averaged vertical tracer profiles via a 1D diffusion equation with an effective 
diffusion coefficient and/or (2) to determine the diffusive flux and resulting $K_{zz}$ profile that best matches the 
horizontally averaged vertical flux in the GCMs.  These latter methods have been shown to result in $K_{zz}$ values 
significantly lower than those obtained by representing $K_{zz}$ as the root-mean-square vertical velocity times the scale height 
\citep{2013A&A...558A..91P,2014A&A...564A..73A}. The exact value of the eddy diffusion coefficient, especially at pressures 
greater than $\sim$0.1 bar, has a direct impact on the location of the quench level of each species (the higher the $K_{zz}$, the 
deeper the quench level) and the abundance at which each species gets quenched in the upper atmosphere.  If the quench point 
falls within the radiative region of the atmosphere, $K_{zz}$ is expected to vary with the inverse square root of atmospheric 
pressure near this quench point \citep{lindzen81,2013A&A...558A..91P}.  If the quench point falls within the deeper 
convective region of the atmosphere, mixing-length theory and free-convection theory can be used to estimate the magnitude 
of $K_{zz}$ near the quench point \citep{visscher10co}, or expressions based on laboratory studies of turbulent rotating 
convection can be developed \citep{wang15}. However, the appropriate mixing length to use for the $K_{zz}$ estimates is not straightforward \citep[e.g.,][]{smith98}, and observations of quenched species may themselves provide our best means of estimating the magnitude of convective mixing at depth.

\begin{figure*}
\centering
\includegraphics[width=0.9\textwidth]{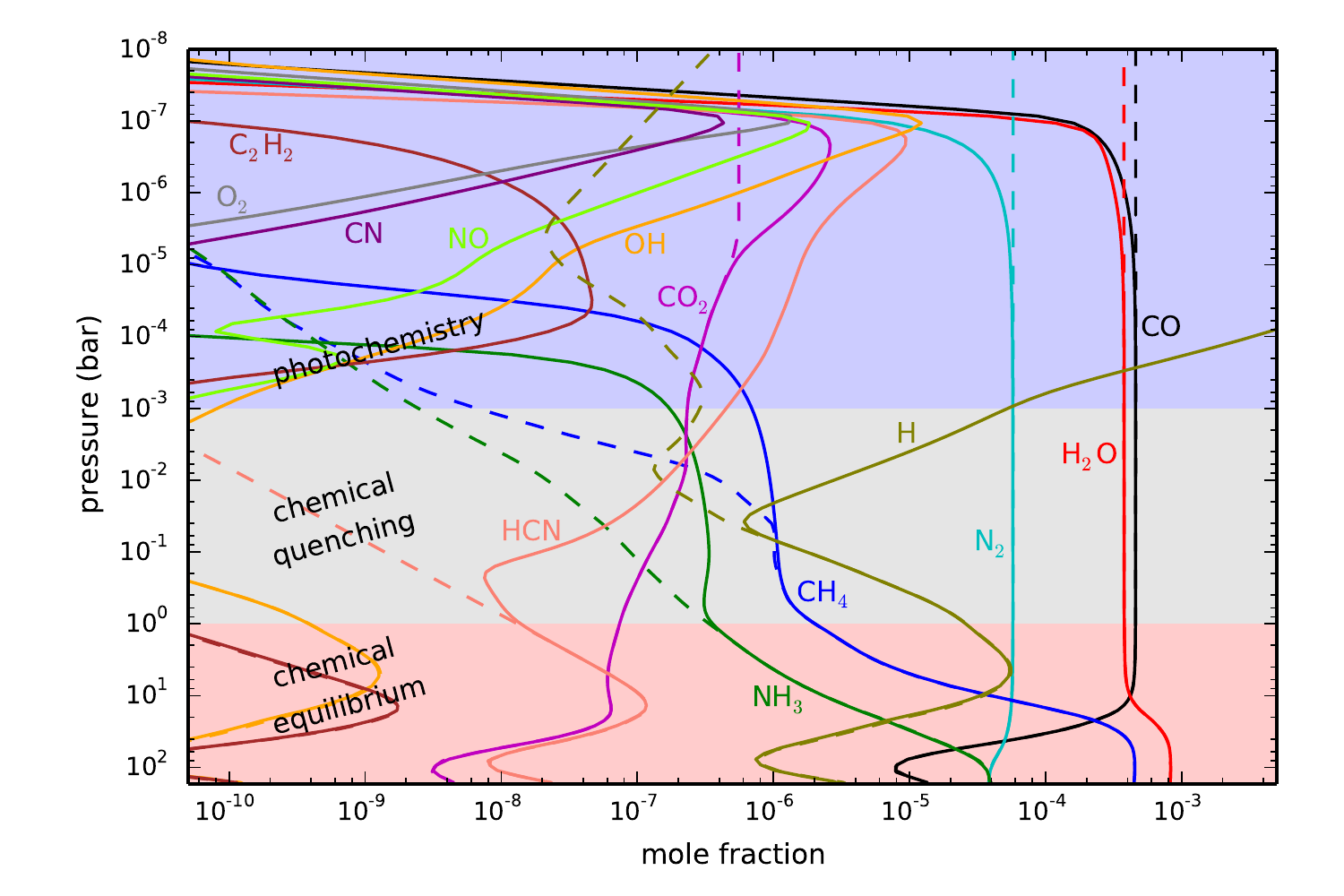}
\caption{Vertical distribution of abundances at the substellar point of HD~189733b. Dashed lines correspond to chemical
equilibrium and solid lines to a one-dimensional vertical model including thermochemical kinetics and photochemistry
\citep{2014A&A...564A..73A}. The atmosphere can be schematically divided into three regions where the composition is 
controlled by either chemical equilibrium, chemical quenching, or photochemistry.}
\label{fig:hd189733b}
\end{figure*}

Chemical quenching in the vertical direction of hot Jupiter atmospheres has been studied through timescale arguments
\citep{2006ApJ...648.1181V,2010ApJ...717..496L,2013MNRAS.435.1888B} and using more robust chemical networks (in which endothermic
reactions are included and reverse reaction rates are computed from thermochemical grounds) suitable to model high temperature
environments
(\citealt{2009arXiv0911.0728Z,2009ApJ...701L..20Z,2011ApJ...737...15M,moses2013a,2012A&A...546A..43V,2012ApJ...745...77K,2014A&A...564A..73A,2014ApJ...780..166M};
see also Fig.~\ref{fig:hd189733b}). Most of these studies focus on the atmospheres of the widely observed hot Jupiters HD~189733b
and HD~209458b, where it is predicted that, assuming nearly solar elemental abundances, CO and N$_2$ are the major reservoirs of
carbon and nitrogen, respectively, H$_2$O is very abundant (it takes up most of the excess of oxygen not locked into CO), and
methane and ammonia are present at a lower level, with CO/CH$_4$ and N$_2$/NH$_3$ abundance ratios in the ranges 50-500 and
5-200, respectively, in HD~189733b, and in the ranges 1000-10,000 and 100-1000, respectively, in the hotter HD~209458b
\citep{2011ApJ...737...15M,2012A&A...546A..43V,2014A&A...564A..73A,2014ApJ...780..166M}. The abundance of carbon dioxide is found
to be somewhat low for atmospheres with an assumed solar elemental composition (see Fig.~\ref{fig:hd189733b}). It is interesting
to note that when CO is the dominant carbon component, CO$_2$ is much less affected by chemical quenching than many other
molecules  because the various reactions involved in the interconversion between H$_2$O-CO-CO$_2$ are rapid enough 
\citep{2011ApJ...737...15M,2012A&A...546A..43V}. The chemistry of elements other than C, N, and O, although more difficult to 
constrain through observations and more uncertain due to a lack of robust reaction rate measurements under appropriate conditions, 
has been also studied theoretically. It has been pointed out, for example, that chemical quenching leads to PH$_3$ becoming a 
major reservoir of phosphorus in hot Jupiter atmospheres \citep{2006ApJ...648.1181V}.

\begin{figure*}
\centering
\includegraphics[width=0.9\textwidth]{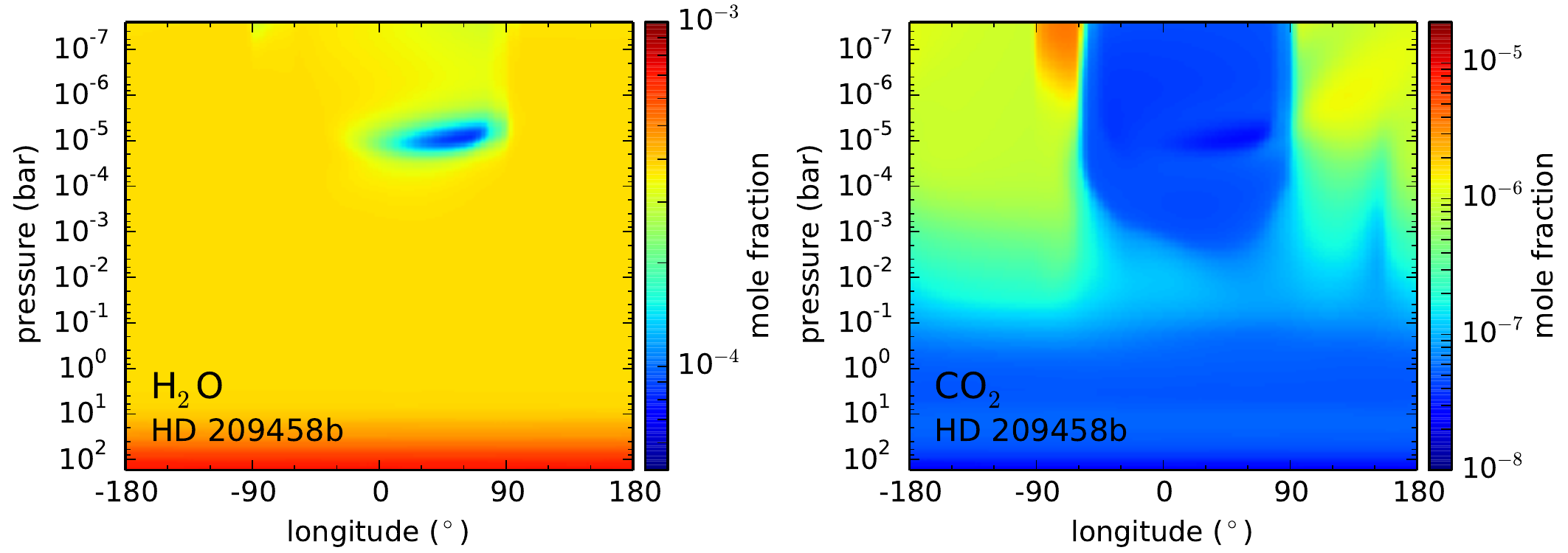}
\caption{Calculated distributions of H$_2$O and CO$_2$ as a function of longitude and pressure in the atmosphere of HD 209458b, from \citet{2014A&A...564A..73A}. Note that H$_2$O maintains a rather homogeneous distribution while CO$_2$ experiences important abundance variations with longitude and height.}
\label{fig:pseudo-2d}
\end{figure*}

Chemical quenching can also occur horizontally driven by winds that move material from hot to cooler atmospheric regions. In hot
Jupiters, which are usually tidally locked (with permanent day and night sides), the uneven heating of the planet may result in
important temperature contrasts, and thus in possible variations in the composition, between the day and night sides. On the
other hand, circulation dominated by a strong superrotating equatorial jet tends to homogenize both the temperature and chemical
composition between the different planetary sides \citep{2002A&A...385..166S,2006ApJ...649.1048C,2007Natur.447..183K}. The main effect of horizontal chemical quenching is that molecular abundances are quenched horizontally to values typical of the hottest dayside regions,
making the cooler nightside to be highly contaminated by the warmer dayside regions
\citep{2006ApJ...649.1048C,2012A&A...548A..73A,2014A&A...564A..73A}. The effect is however more marked for molecules such as
H$_2$O, CO, and N$_2$, which show rather uniform abundances, than for others such as CH$_4$, NH$_3$, CO$_2$, and HCN, which may
still show important abundance gradients among the different planetary sides (see Fig.~\ref{fig:pseudo-2d}). The horizontal homogenization of abundances becomes more important as the atmosphere gets cooler. The distribution of the atmospheric constituents has implications for transit observations, which probe the terminator region, and for phase curves probing the different planetary sides.

\subsection{Photochemistry}
\label{subsec:photochemistry}

Hot Jupiters are strongly irradiated by their host stars and thus receive a high ultraviolet flux, which is absorbed in the upper
atmosphere and induces a more or less rich photochemistry. That is, the parent molecules transported from deep atmospheric regions are photodissociated in upper layers producing radicals that react to form new species. The study of the photochemistry of exoplanet atmospheres has benefited from decades of study of solar system atmospheres such as those of Jupiter \citep{1996Icar..119....1G}, Saturn \citep{2000Icar..143..244M}, and Titan \citep{1984ApJS...55..465Y,1996JGR...10123261L}. However, some of the chemical kinetics and photoabsorption cross section data still need to be adapted to the high temperatures of hot Jupiters \citep{2013A&A...551A.131V}. The type of molecules formed by photochemistry depends on the composition of the precursor material transported from deeper layers, which in turn depends on the elemental composition (see section~\ref{subsec:metallicity}) and thermal atmospheric structure (see sections~\ref{subsec:equilibrium} and \ref{subsec:quenching}). The cooler the planet the larger the extent of photochemistry because, on the one hand, high temperature chemistry counterbalances the action of photochemistry, and on the other, low temperatures favour the presence of CH$_4$ and NH$_3$, which are more active photochemical precursors than their respective high-temperature counterparts CO and N$_2$ \citep{2014RSPTA.37230073M}.

There are various species produced by photochemistry in the atmospheres of gas giant planets (see Fig.~\ref{fig:hd189733b}).
These atmospheres being hydrogen-rich, a major species produced by photochemistry is atomic hydrogen, which is formed by
photolysis and thermal decomposition of H$_2$ as well as by catalytic photolysis of H$_2$O \citep{2003ApJ...596L.247L}. The
higher the irradiation of the planet the more abundant becomes H. Another important photochemical species is HCN, which is efficiently formed in the photochemical layer through schemes starting with the photodissociation of CH$_4$ and NH$_3$ (provided the atmosphere is not too hot to exclude these two hydrides as important constituents), and is able to diffuse to deeper regions \citep{2009arXiv0911.0728Z,2011ApJ...737...15M}. The photodissociation of CH$_4$ also triggers the formation of larger hydrocarbons such as C$_2$H$_2$, which may then polymerize to form soots \citep{2009arXiv0911.0728Z,2013MNRAS.435.1888B}. Another important photochemical product is the radical OH, which results from the photodissociation of water and act as a key intermediate in the synthesis of other O-bearing molecules such as O$_2$ and NO \citep{2011ApJ...737...15M}. 
As concerns sulfur compounds, the major photochemical products of the dominant equilibrium sulfur species, H$_2$S, are expected to be S, HS, S$_2$, SO, and SO$_2$ \citep{2009ApJ...701L..20Z}.   
\citet{lavvas14alkali} show that for planets hot enough that refractory elements are not tied up in condensates, atomic 
neutrals and ions such as Mg, Mg$^+$, Fe, Fe$^+$, Ca, Ca$^+$, Na, Na$^+$, K, K$^+$, Si$^+$, Al$^+$ could be important 
photochemical and equilibrium species, although some dominant equilibrium molecular forms (e.g., 
SiO, SiS, AlOH, FeH) can also survive in the photosphere on hot planets \citep[see also][]{visscher10rock}, and molecular 
species like NaCl and KCl may form from both photochemistry and thermochemistry.

On highly irradiated hot Jupiters, most photochemical products are confined to relatively high altitudes (e.g., above $\sim$0.1
mbar, see Fig.~\ref{fig:hd189733b}). Therefore, photochemical products are likely to have a limited impact on the infrared 
emission spectrum of the 
planet, whose photosphere is located at deeper layers, although they may leave their imprints on the transmission spectrum, 
which can probe higher atmospheric layers.  The cooler the planet, however, the deeper the photochemical ``layers'' extend, and the
higher the likelihood that the photochemical products affect the infrared emission spectra.  On our own solar-system giant planets, 
for example, methane photochemical products such as C$_2$H$_2$ and C$_2$H$_6$ survive throughout the stratosphere, dominating the 
mid-IR spectra and enabling more effective cooling of the stratospheres of these planets.  In the case of HCN, its extended 
vertical distribution and relatively high abundance in cool to moderately warm gas giant planets makes it a very likely 
candidate for detection \citep[e.g.,][]{2011ApJ...737...15M}.

One-dimensional models indicate that in still upper layers, molecules are completely photodestroyed and atoms become the 
major constituents.  In the absence of a hydrodynamic escape flow, molecular diffusion at a high-altitude homopause layer 
(typically in the 10$^{-6}$ to 10$^{-9}$ bar range for hot Jupiters) will cause the atmospheric constituents to become 
layered according to their weight, following their own density scale heights.  Because of both this molecular diffusion and 
the strong X-ray/ultraviolet irradiation from the host star (which strongly heats the high-altitude thermosphere), atomic 
hydrogen becomes the only remaining dominant neutral constituent at high altitudes on close-in giant planets 
\citep[e.g.,][]{2011ApJ...737...15M}.  If the upper atmosphere were to remain cool, the molecular region could extend to 
very high altitudes \citep[e.g.,][]{2014A&A...564A..73A}.  However, thermospheric photochemical models that include X-ray and
extreme ultraviolet photolysis and photodissociative ionization of H$_2$, ion chemistry, and realistic atmospheric escape 
processes \citep{2007P&SS...55.1426G,2013Icar..226.1678K} demonstrate that hydrodynamic winds fueled by the escaping gas 
can drag heavy atoms and ions to regions as high as the exosphere in strongly irradiated planets, but that molecules 
(including H$_2$) cannot survive the large thermospheric temperatures on such planets.
In fact, various types of atoms, from the lightest hydrogen to heavy metals, have been detected in the vertically extended, 
escaping atmospheres of hot Jupiters
\citep{vidal-madjar2003,vidal-madjar2004,vidal-madjar2013,etangs2010,fossati2010,linsky2010}.

Ionization of atmospheric constituents by galactic cosmic rays provides an additional source of disequilibrium 
chemistry on exoplanets. \citet{rimmer13} and \citet{rimmer14} have explored the effects of cosmic rays on 
the atmospheric compositions of extrasolar giants planets.  Charging of mineral grains, and potential subsequent 
electrical discharges, are another possible source of disequilibrium chemistry on extrasolar giant planets 
\citep[see][]{helling13part3,bailey14,stark14}.

\subsection{Sensitivity to bulk elemental composition}
\label{subsec:metallicity}

Gas giant planets are expected to have nearly solar elemental compositions. However some may show a different metallicity, i.e., the abundances of elements heavier than helium scaled up or down by a similar factor, and/or elemental abundance ratios different from those found in the Sun. These deviations from the solar elemental composition, whatever their origin (see section 6), can have a significant impact on the atmospheric composition.

In general, an enhancement of metallicity in H/He-dominated atmospheres of gas giant planets favors an increase in the
abundances of molecules that contain multiple heavy atoms, i.e., CO and N$_2$ are favored over CH$_4$ and NH$_3$.
Molecules with more than two heavy atoms are even more favored, as occurs in the case of CO$_2$, whose abundance
increases as the square of metallicity \citep{2009arXiv0911.0728Z,2009ApJ...701L..20Z}. See more details on this subject
in sections \ref{subsec:equilibrium_nep} \& \ref{subsec:disequilibrium_nep}.

In warm H/He-dominated atmospheres, one of the most critical elemental ratios is C/O. The issue is well known in the study of
evolved stars, some of which can bring out to the surface enough carbon to revert the C/O abundance ratio, which in the Sun is
0.55 \citep{asplund2009}. The high stability of CO under these conditions makes it lock almost all the limiting reactant,
either C or O, allowing for the reactant in excess to form O-bearing molecules such as H$_2$O when C/O $<$ 1 and C-bearing
molecules such as HCN and C$_2$H$_2$ when C/O $>$ 1 \citep{1973A&A....23..411T}. The observational claim of a carbon-rich
atmosphere in the hot Jupiter WASP-12b \citep{madhusudhan2011b}, although still subjected to debate 
\citep{cowan2012,crossfield2012,madhusudhan2012,swain2013,mandell2013,stevenson2014a,stevenson2014b,kreidberg2015},
has opened a window on the possible existence of carbon-rich giant extrasolar planets. Whatever their origin (see section 6),
such planets would show a chemical composition dramatically different from the more traditional gas giants with a nearly solar
C/O ratio (i.e., oxygen-rich). In moderately warm and hot carbon-rich atmospheres, water is no longer an abundant constituent,
the C-bearing molecules C$_2$H$_2$, CH$_4$, and HCN become major constituents, and CO$_2$ vanishes to a negligible level
\citep{madhusudhan2012,2012ApJ...745...77K,moses2013a,moses13gj436b,2015arXiv150203567V,heng2015}. 
If the atmosphere is cool enough ($<$ 1000 K) then CH$_4$ and H$_2$O become the main reservoirs of carbon and 
oxygen, respectively, at the expense of other carbon- and oxygen-bearing molecules \citep{madhusudhan2012,moses2013a,moses13gj436b,2015arXiv150203567V}. 

\subsection{Clouds and hazes}
\label{subsec:clouds}

The terms \emph{cloud} and \emph{haze} are often used interchangeably, but here we follow the spirit of \cite{marley2013} and use these terms to refer to two fundamentally different types of condensates that may appear in the atmospheres of gas giant planets. 
By \emph{cloud} we refer to the typical ``cooling clouds'' \citep[e.g.,][]{rossow78} that form when a volume of gas is cooled
dynamically or radiatively, such that the partial pressure of a constituent or constituents exceeds its saturation vapor
pressure and condensation ensues.  The typical equilibrium condensates expected along the cloud-condensation sequence in a 
hydrogen-dominated atmosphere \citep[e.g.,][]{lodders10,marley2013,morley2013} are thus considered clouds, regardless of 
their optical depth, horizontal extent, etc.  We restrict the term ``hazes'' to refer to aerosols that form \emph{in situ} by 
the action of photochemistry or any other disequilibrium chemical process (e.g., the so-called ``soots'' discussed by 
\citealt{2009arXiv0911.0728Z}) or from photochemically produced gases flowing dynamically into cooler regions, where they 
can condense.

In the rainout scenario of cloud formation, a condensate forms and gravitationally settles in the atmospheric layer where the temperature equals its condensation temperature, and the atmosphere above the cloud becomes depleted in the elements that take part in that condensate \citep{1969Icar...10..393L,1994Icar..110..117F,1996ApJ...472L..37F,1999ApJ...512..843B,1999ApJ...519..793L,2002Icar..155..393L,sudarsky2003}.  Thus, as one moves from the deep and hot atmosphere to upper and cooler regions, elements are progressively removed from the gas phase according to their refractory character. Chemical equilibrium calculations including gas and condensed species are very useful to identify the most plausible cloud-forming species (see Fig.~\ref{fig:pt_diagram}). In the rainout approach, chemical equilibrium is solved in order of decreasing temperature and, when a given species is found to condense the elements which take part in that condensate are removed (in the corresponding stoichiometric proportions) before continuing to solve chemical equilibrium at lower temperatures. Thus, it is not only necessary to know the possible condensates but also the condensation sequence and the main condensates that deplete each element.

Unlike the case of Jupiter, whose atmosphere is so cold that water and ammonia condense to form tropospheric clouds, most
extrasolar giant planets characterized to date are sufficiently hot to ensure the survival of water vapor and other volatiles 
in their atmospheres. In hot Jupiters only the most refractory species can condense to form clouds.  Although the exact 
temperature-pressure profile of the planet in question controls which refractory species will condense first (i.e., at the 
deepest pressures; see \citealt{lodders10}), aluminum, titanium, and calcium are expected to be the first elements to be 
removed from the gas phase, condensing as corundum (Al$_2$O$_3$), Ca-aluminates such as hibonite (CaAl$_{12}$O$_{19}$), and 
Ca-titanates such as perovskite (CaTiO$_3$). 
Therefore, atmospheres hotter than about 2000 K can maintain titanium in the gas phase to form oxides such as TiO, while cooler
atmospheres would deplete most of this element in the form of perovskite or other Ca-titanate \citep{1999ApJ...512..843B,2002ApJ...577..974L}. In fact, it has been proposed that in the atmospheres of very hot giant exoplanets, the survival in the gas phase of species such as TiO and VO can provide a sufficiently high opacity at optical wavelengths as to induce a thermal inversion \citep{2003ApJ...594.1011H,2008ApJ...678.1419F,2009ApJ...699.1487S}. The detections of TiO and VO in brown dwarf atmospheres at high temperatures \citep{kirkpatrick2005} provide support to this proposition. However, TiO and VO have not yet been unambiguously identified in the atmospheres of strongly irradiated hot Jupiters \citep{2008A&A...492..585D,huitson2013,2013MNRAS.436....2M,swain2013,sing2013,2013MNRAS.436.2974G,stevenson2014b,2014ApJ...783....5S, 2015A&A...575A..20H}, though recently a potential detection of TiO has been reported for one of the most irradiated hot Jupiters WASP-33b \citep{haynes2015}. 

At temperatures below $\sim$1800 K, iron is expected to condense homogeneously, and magnesium, together with a good fraction of silicon, are expected to deplete in the form of silicates such as Mg$_2$SiO$_4$ and MgSiO$_3$. The presence of such type of clouds has been inferred from transit observations of the hot Jupiter HD~189733b \citep{pont2008,pont2013,2008A&A...481L..83L,2009A&A...505..891S,gibson2012}. The alkali metals Na and K can be depleted in the form of the aluminosilicates NaAlSi$_3$O$_8$ and KAlSi$_3$O$_8$ at temperatures around 1500 K \citep{1999ApJ...512..843B}, although in the rainout scenario it is likely that at these temperatures most of the aluminum is no longer available in the gas phase, in which case Na and K would condense at $\sim$1000 K as Na$_2$S and KCl, respectively \citep{1999ApJ...519..793L}. In sufficiently hot atmospheres neutral Na and K atoms can survive in the gas phase and can be readily detected through the Na~{\small I} doublet at 589.0 nm and 589.6 nm and the K~{\small I} doublet at 766.5 nm and 769.9 nm \citep{charbonneau2002,redfield2008,wood2011,sing2011a,2012MNRAS.426.1663S,nikolov2014,2014A&A...563A..41M}.

The rainout scenario provides a useful methodology to predict the composition of the clouds that may be present in a given
atmosphere and the base level at which each cloud forms. However, the computation of the size and concentration of particles
above this level and the horizontal distribution of the clouds, i.e., how particles nucleate and grow and are affected by 
atmospheric dynamics, still remains a formidable challenge. Nevertheless, efforts to this end have been undertaken by various groups \citep{2001ApJ...556..872A,2008MNRAS.391.1854H,2010A&A...513A..19F,2013A&A...558A..91P,morley2013,lee2015}.

The photochemical formation of organic hazes in the atmospheres of hot Jupiters has been addressed in a couple of theoretical
studies \citep{2004ApJ...605L..61L,2009arXiv0911.0728Z}. The general view provided by these studies is that in hot atmospheres,
most of the carbon remains locked into CO, while in cooler atmospheres the larger amount of carbon stored into CH$_4$ can be
photochemically driven to larger hydrocarbons, eventually producing hydrocarbon aerosols or soots. Cooler planets would therefore
be expected to have more prevalent photochemical hazes.  This scenario agrees with the
presence of hazes in the cold atmospheres of Jupiter, Saturn, and Titan (e.g.,
\citealt{2005Natur.438..765T,2013Icar..226..159Z}), although it is in contradiction with the inference of aerosols in
the very hot atmosphere of WASP-12b \citep{sing2013}, unless these aerosols result from the most refractory
species along the cloud-condensation sequence (e.g., Al$_2$O$_3$) at the cooler limb of the planet \citep{sing2013}, 
or unless the planet is indeed carbon-rich (and thus contains the more photochemically active molecules HCN and C$_2$H$_2$, e.g.,
\citealt{2012ApJ...745...77K} and \citealt{moses2013a}). It is clear that more work is needed in both the theoretical 
and observational sides to better understand the formation and role of hazes in extrasolar giant planets.

\subsection{Hot Jupiters versus directly imaged planets}
\label{subsec:imaged}

One of the most recent and formidable successes achieved in the field of exoplanets has been the detection by direct imaging of
young and self-luminous gas giant planets, opening the way to characterize their atmospheres by direct spectroscopy
\citep{2010ApJ...710L..35J,janson2013,2010ApJ...716..417H, bowler2010,2011ApJ...729..128C,barman2011a,2012ApJ...753...14S,oppenheimer2013,konopacky2013,2015ApJ...798L...3C}.
The few planets that have been characterized by this method are more massive than Jupiter and have effective temperatures in the
range 600-1700 K. Thus, they have some similarities with brown dwarfs and free-floating planets
\citep{2000Sci...290..103Z,2001RvMP...73..719B,burrows2003}. Directly imaged planets share also some characteristics with hot Jupiters, as both are gas giants and hot. However, unlike hot Jupiters, directly imaged planets orbit far from their host star and are young, so that they are heated predominantly from the interior rather than irradiated by the star. This difference leads to qualitative differences in the dynamical and thermal structure of the atmosphere. In hot Jupiters, the high irradiation causes the atmospheres to be radiative in almost the entire observable atmosphere, and the temperature profile is isothermal in the lower atmosphere, at pressures above $\sim$1 bar \citep[see e.g.][]{burrows2008}, before convection dominates in the deep atmosphere ($\gtrsim$100 bar). On the other hand, the atmospheres of self-luminous planets and brown dwarfs are largely driven by convection, leading to adiabatic temperature gradients even in the observable atmospheres. 

The atmospheres of hot Jupiters and directly imaged planets have temperatures of the same order and are expected to have a nearly
solar elemental composition, so that one would expect a similar atmospheric chemistry in both types of planets. There are,
however, a couple of major differences. First, the atmospheres of directly imaged planets are likely less affected by
photochemistry than in the case of hot Jupiters because of their much larger orbital distances; however, directly imaged planets
tend to orbit young stars, and young stars tend to have high ultraviolet output, so photochemistry will not be negligible on
these planets. And second, the source of heat being located in the interior of the planet rather than outside imprints
differences in the atmospheric thermal profile. According to the recent theoretical study by \cite{2014ApJ...797...41Z}, the
higher temperatures in the deep atmosphere of self-luminous planets favor CO over CH$_4$ in deep layers, but also in the upper
observable atmosphere as a consequence of upward mixing. Methane is therefore predicted to be a minor atmospheric constituent in
most self-luminous planets characterized to date, except for a couple of planets with cool atmospheres whose near-infrared spectra show evidence for CH$_4$ absorption, such as GJ~504b \citep{2013ApJ...778L...4J}, 51 Eri b \citep{macintosh2015}, and HR~8799b \citep{barman2015}. Years of study of brown dwarf atmospheres \citep[see e.g.,][]{marley2015}, which share many similarities with the atmospheres of self-luminous planets, provide an invaluable basis to aid in the understanding of atmospheric chemistry of self-luminous planets, although further observations are still needed. 

\section{Theory of Atmospheric Chemistry of Exo-Neptunes and Super-Earths}
\label{sec:chemistry_nep}

The same chemical processes discussed above for giant planets --- thermochemical equilibrium, disequilibrium 
quenching due to transport, and disequilibrium photochemistry --- also affect smaller planets, but the basic 
ingredients available to the atmospheres of smaller planets differ from those of gas giants.  These differences 
emerge early on in the planet's evolution, with such factors as the dust/planetesimal surface density 
distribution within the protoplanetary disk (see chapters on protoplanetary disk evolution in present issue, Oberg et al. 2016 and Fang et al. 2016), the planet's initial formation and feeding-zone location within 
the disk, and the planet's migration history being important parameters that control how massive the planet 
becomes, what atmospheric volatiles are collected, and how much atmospheric hydrogen and helium are retained 
\citep[e.g.,][]{pollack1996,lissauer07,dangelo10,dodson10,marboeuf2014,ali-dib2014,madhusudhan2014c,helling2014}.  
In the gravitational-instability theory for giant-planet formation \citep[e.g.,][]{boss97}, giant planets 
are expected to end up with a metallicity similar to the host star.  In the core-accretion theory of giant-planet formation 
\citep[e.g.][]{pollack1996}, protoplanetary cores form from the accretion of solid planetesimals, and gas 
accretion rates are initially slow.  As the solid protoplanetary core grows, it can accrete more and more of 
the surrounding hydrogen-rich nebular gas \citep[][also see chapter by Massol et al. 2016 in present issue]{lammer2014,stokl2015}.  When the core becomes massive enough, with a critical mass of 
order $\sim$10 Earth masses \citep[but cf.][]{Venturini2015}, the protoplanet can experience a runaway gas-accretion phase in which the nebular gas is rapidly accreted onto the protoplanetary core.  Planets that reach this runaway gas-accretion 
stage become hydrogen- and helium-rich gas-giant planets.  Planets that don't reach this stage contain 
less H and He.

Whether a protoplanet can reach this runaway gas-accretion stage or not depends largely on the accretion rate 
of solids in comparison to the time scale for dissipation of gas from the disk.  Just beyond condensation fronts 
in the disk, such as the water-ice ``snow line'', the surface density of solid material becomes large, leading 
to more rapid accretion of solids and a reduced time scale for formation of massive protoplanetary cores 
\citep[e.g.,][]{stevenson88,ciesla06,dodson09}. The formation of giant planets is thus expected to be 
particularly efficient near the disk snow line.  In lower-surface-density regions of the disk farther out 
from the snow line, the solid accretion rate is slower, and a protoplanet may never reach this runaway 
gas-accretion phase, leading to less accumulation of the nebular gas.  In this classical picture, the 
formation of Uranus and Neptune occurred too slowly to allow these planets to fully reach the runaway 
gas-accretion phase, leading to ``ice giant'' planets whose total mass is dominated by heavier volatiles 
such H$_2$O rather than hydrogen and helium \citep[e.g.,][]{lissauer07,fortney10ssr}.  The overall initial 
mass fraction of H/He in a planet's atmosphere is a sensitive function of the disk characteristics and 
lifetime, with small changes in disk properties or evolutionary history having a significant impact on 
the resulting mass and composition of planets that are intermediate in size between gas giants and solid 
terrestrial planets \citep[e.g.,][]{helled14,lammer2014,luger2015}. While generally atmospheres of ice-giants are expected to be H$_2$-rich, recent theoretical studies have also suggested the possibility of He-dominated atmospheres \citep{hu2015b}.  

In this section, we discuss the atmospheric chemistry of these intermediate-sized planets, the so-called 
super-Earths, mini-Neptunes, and Neptune-sized planets, with radii spanning $\sim$1 -- 6 $R_{\oplus}$, 
that are observed to constitute a large percentage of the known planets in our 
galaxy \citep[e.g.,][]{borucki11,mayor11,Howard2012,batalha13,Fressin2013,burke14,rowe15}.  These planets will  
typically contain less hydrogen and helium (and thus a higher atmospheric metallicity) than giant planets 
because of the less efficient accretion of nebular gas during their formation and evolution, as well as 
the higher likelihood of the escape of light gases over time.  Outgassing from the planetary interior is 
expected to contribute additional volatiles --- a component of the atmosphere that will become increasingly 
important for smaller planets --- and other evolutionary processes such as atmospheric escape, impact 
delivery or erosion, atmosphere-surface exchange, weathering, and sequesteration of volatiles into the 
interior can have and major influence on atmospheric composition and chemistry.  The stochastic nature of 
the different possible evolutionary pathways is expected to lead to highly diverse atmospheric properties 
for intermediate and small planets 
\citep[e.g.][]{pepin06,dodson10,rogers11,lopez13,fortney13,2013AREPS..41..469H,leconte14}.  Here, we consider 
the atmospheric chemistry of exo-Neptunes and super-Earths with widely diverse volatile contents, as well as 
the chemistry of outgassed atmospheres of smaller, hot, rocky planets. The atmospheric chemistry of 
terrestrial planets near the habitable zone is briefly discussed in section~\ref{sec:habitability}. 

\subsection{Chemical Equilibrium in Exo-Neptune and Super-Earth Atmospheres}
\label{subsec:equilibrium_nep}

Thermochemical equilibrium can be maintained in high-temperature, high-pressure regions of exoplanet 
atmospheres (see section~\ref{subsec:equilibrium}), so equilibrium conditions are appropriate to consider 
to first order for any super-Earths or exo-Neptunes with thick, hot atmospheres.  This statement is true 
whether the planet is a low-density fluid ``ice giant'', like Neptune itself, or has a thick atmosphere 
overlying a solid surface, like Venus.  Even secondary atmospheres produced from the outgassing of interior 
volatiles during and after the accretion phase can lead to thick, high-pressure atmospheres that remain hot 
at depth due to accretional energy, radioactive decay in the interior, tidal heating, strong 
stellar irradiation, and/or a greenhouse effect.  

Our own terrestrial planets have demonstrated the importance of secondary outgassing of interior 
volatiles in shaping the atmospheric properties of solid-surface planets \citep{pepin06}.  The 
theoretical equilibrium composition of super-Earth atmospheres dominated by such an outgassing source 
has been explored by several investigators.  For example, \citet{elkins08} have examined the wide range of 
atmospheric masses and compositions that can result from the degassing of different meteoritic 
compositions during the planet's accretion phase.  \citet{schaefer10} have performed similar more detailed 
calculations, albeit with a focus on the early Earth, to investigate the chemistry of potential steam 
atmospheres, such as those predicted to be associated with magma oceans \citep[e.g.,][]{abe85} thought 
to form during the accretion of the Earth and other terrestrial (exo-)planets.  Both \citet{schaefer10} 
and \citet{elkins08} find that the resulting atmospheric composition is a sensitive function of the 
assumed composition of the meteoritic material being accreted --- water-dominated steam atmospheres 
occur only for CI and CM chondritic starting material.  For other assumed meteoritic starting compositions, 
the atmospheres can be dominated by CO$_2$, N$_2$, H$_2$, CH$_4$, or CO, depending on the starting material 
composition and atmospheric temperatures \citep{schaefer10}.  \citet{schaefer11} have also pursued the 
potential atmosphere-surface buffering of hot atmospheres in equilibrium with planetary surfaces.

The possible formation of exotic (by solar-system standards) silicate atmospheres on hot super-Earths, 
in which volatile elements such as H, C, N, S, and Cl have already escaped from the planet, is explored 
by \citet{schaefer09} and \citet{miguel11}.  In these calculations, the atmosphere is assumed to be 
in gas-melt equilibrium with a volatile-free magma ocean or partially molten lithosphere.  Their results indicate 
that such atmospheres can be composed largely of atomic Na, O$_2$, O, with SiO, Fe, and/or Mg, depending on the 
planet's orbital distance and resulting atmospheric temperature.  \citet{ito15} have performed a similar 
set of equilibrium calculations for super-Earth atmospheric compositions over a volatile-free magma ocean 
of various assumed compositions; they discuss the resulting thermal structure, spectroscopy, and detectability 
of such atmospheres.  They find that SiO, in particular, affects the atmospheric opacity, causing thermal 
inversions and notable infrared emission signatures.

\begin{figure*}[hb]
\centering
\includegraphics[width=0.9\textwidth]{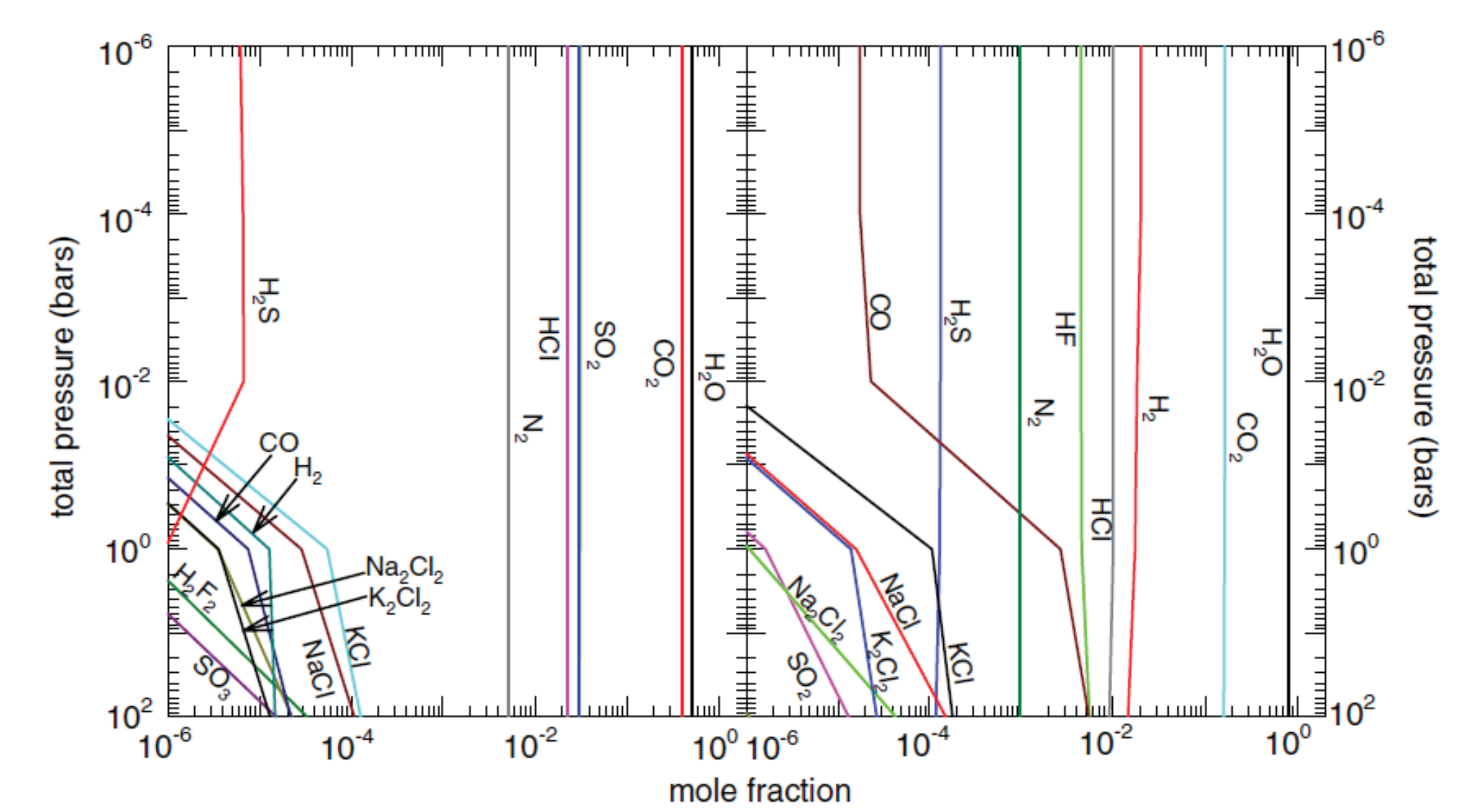}
\caption{Equilibrium atmospheric composition for a GJ 1214b-like exoplanet, assuming the atmosphere 
results from outgassing from a high-temperature felsic silicate terrestrial crustal composition (left) 
or a more mafic bulk-Earth silicate composition (right).  Figure from \citet{schaefer12}.}
\label{fig:schaeferfig}
\end{figure*}

\citet{schaefer12} have extended their earlier 2010 investigation to consider a wider range of parameter 
space that might be relevant to outgassed atmospheres of hot terrestrial exoplanets and super-Earths, 
including considerations of variations in the relative abundances of elements such as H, C, and O.  Again, 
these authors emphasize that the equilibrium atmospheric composition of such secondary outgassed atmospheres 
depends strongly on the source material and temperatures.  Outgassing from accreted material with the 
composition of the bulk silicate Earth or the terrestrial crust would result in a water-dominated atmosphere 
under conditions relative to GJ 1214b, with CO$_2$ as an important secondary component (see 
Fig.~\ref{fig:schaeferfig}), whereas O$_2$ and Na could dominate in the hotter, lower-density atmosphere of 
CoRoT-7b (\citealt{schaefer12}; see also \citealt{leger11}).  The chemistry of possible hot equilibrium 
atmospheres of Earth-like planets after giant impact events is explored by \citet{lupu14}, who find that 
the main consituents are water and CO$_2$, along with smaller amounts of HCl, H$_2$, HF, CO, N$_2$, alkali 
halides, SO$_2$, and H$_2$S.  

The equilibrium atmospheric composition of warm super-Earths like GJ 581c is investigated by \citet{miller09}, 
for an assumed cometary-like complement of volatiles with variable hydrogen content.  For hydrogen-rich 
atmospheres, the dominant constituents are H$_2$, H$_2$O, CH$_4$, and NH$_3$.  For the hydrogen-poor situation, 
the resulting Venus-like atmosphere has dominant constituents CO$_2$ and N$_2$, with much less H$_2$O.  For the 
``intermediate'' hydrogen case, the atmosphere has an interesting mixture of dominant constituents like 
H$_2$O, H$_2$, CO$_2$, and CH$_4$.  The overall hydrogen content therefore has a strong influence on the 
resulting composition.

Observations of exoplanets for which both mass and radius have been determined show a transition in 
planetary bulk densities near radii of $\sim$1.5--2 $R_{\oplus}$; the smallest planets tend to be dense and 
presumably rocky, with at best a tiny mass fraction of light volatiles like hydrogen and helium, and larger 
planets tend to require increasingly larger mass fractions of H/He envelopes to explain the bulk density
\citep[cf.][]{weiss13,weiss14,marcy14,wu13,hadden14,rogers15}. Population-synthesis models and other theoretical 
arguments suggest that the trend of decreasing hydrogen and helium content with decreasing planetary size is a 
natural consequence of planetary formation and evolution
\citep{miller11,rogers11,lopez12,fortney13,owen13,lopez13,lopez14,benz14,wolfgang14}.  
Regardless of whether close-in exo-Neptunes and super-Earths with comparatively large H/He contents formed 
farther out in the disk and migrated/scattered inward
\citep[e.g.,][]{ida05,alibert06,terquem07,kennedy08,mordasini12,inamdar15} 
or whether they formed \emph{in situ} \citep{hansen12,hansen13,chiang13,chatterjee14}, the amount of 
hydrogen in the atmospheric envelope has a strong influence on the resulting atmospheric composition.
The metallicity (or bulk mole fraction of H, in general) of the atmosphere is therefore an important parameter 
controlling the chemistry of low-density super-Earths and exo-Neptunes.

\begin{figure*}
\includegraphics[width=\textwidth]{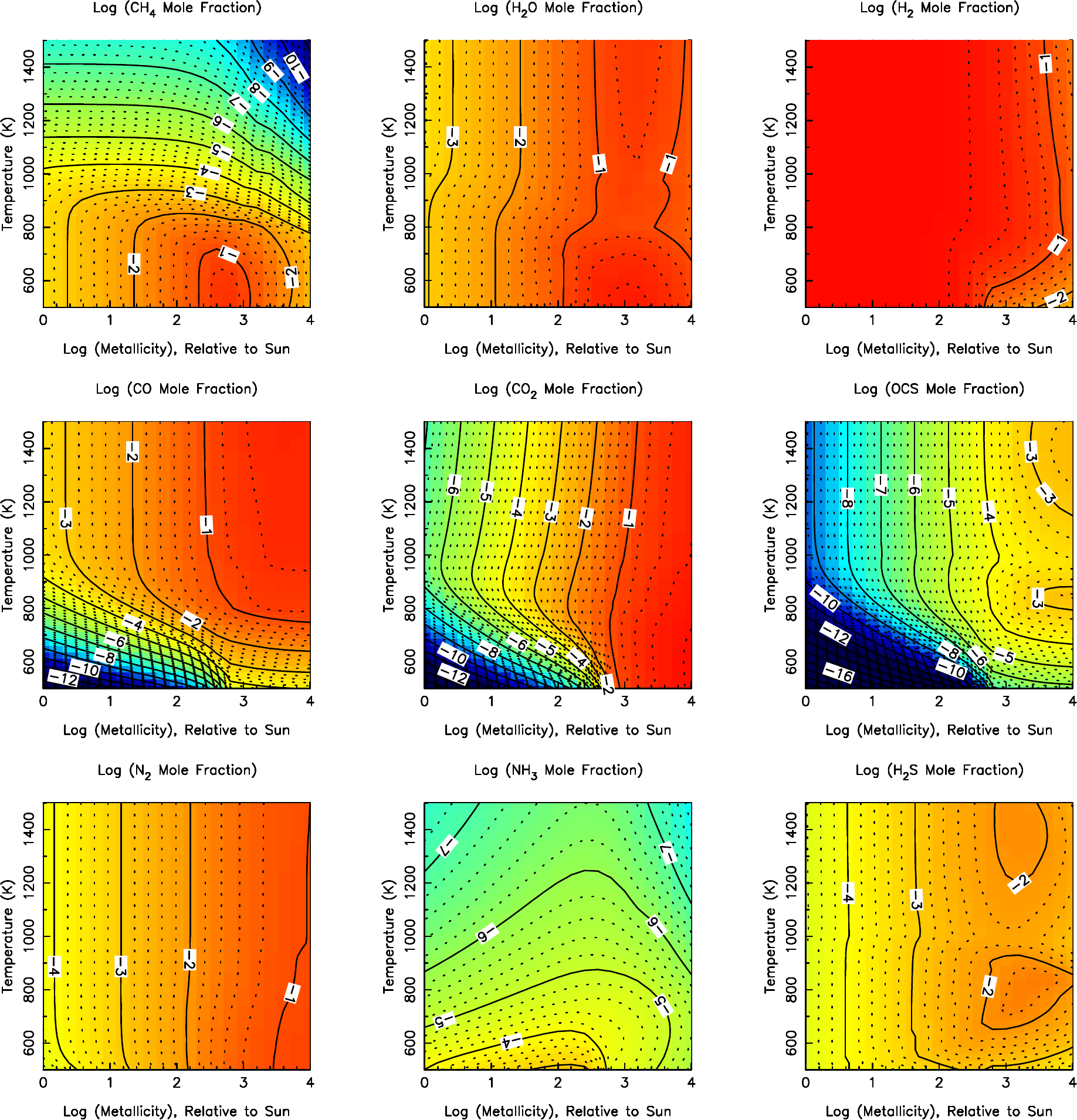}
\caption{Equilibrium mixing ratios for various atmospheric constituents as a function of temperature 
and metallicity for an assumed typical photospheric pressure of 0.1 bar.  Figure adapted from 
\citet{moses13gj436b}.}
\label{fig:nepequil}
\end{figure*}

The effect of metallicity on the equilibrium composition, temperature structure, and/or spectra of 
exoplanet atmospheres has been studied by \citet{2002Icar..155..393L}, \citet{2006ApJ...648.1181V}, 
\citet{visscher10rock}, and \citet{fortney08color} for metallicities up to 3--5$\times$ solar; 
\citet{2009ApJ...701L..20Z,2009arXiv0911.0728Z}, \citet{spiegel10}, \citet{fortney10trans}, 
\citet{2010ApJ...720..344L}, \citet{2011ApJ...738...32L} for metallicities up to 30--50$\times$ 
solar; \citet{2014ApJ...781...68A} and \citet{2014A&A...562A..51V} for metallicities up to 
100$\times$ solar; and \citet{moses13gj436b}, \citet{hu14}, and \citet{miguel15} for very high 
metallicities (e.g., up to and beyond 1000x solar).  These investigations reach concensus on several 
general trends.  First, high-metallicity planets will have higher temperatures at lower pressures 
than otherwise similar low-metallicity planets, due to greater atmospheric opacity from the heavy 
(i.e., non-H$_2$) molecular constituents; that is, the photosphere will shift to higher altitudes.  
Second, the CO/CH$_4$ and N$_2$/NH$_3$ equal-abundance curves shift to lower temperatures with 
higher metallicities, leading to a higher likelihood that CO and N$_2$ will be important carbon 
and nitrogen components, respectively, of a high-metallicity planetary atmosphere.  Third, molecules 
with multiple heavy elements, such as CO, CO$_2$, N$_2$, CS, S$_2$ become favored at the expense of 
molecules that just contain one heavy element, like CH$_4$, NH$_3$, H$_2$S, as the metallicity 
increases --- an effect that is particularly notable for molecules that contain more than two heavy 
elements, such as CO$_2$ and OCS (see Fig.~\ref{fig:nepequil}).  At sufficiently high metallicities, 
CO$_2$ will even replace H$_2$ as the dominant atmospheric constituent for an otherwise solar-composition 
atmosphere.  Molecular hydrogen remains a major constituent of the atmosphere under the conditions 
studied in Fig.~\ref{fig:nepequil}, but it ceases to dominate at the highest metallicities considered 
(e.g., several thousand times solar).  Water increases roughly linearly with metallicity until metallicities 
of $\sim$1000 times solar, at which point the lower H mole fraction begins to adversely impact all 
H-bearing molecules.  The expected depletion of hydrogen in super-Earths and exo-Neptunes due to 
inefficient accretion of the nebular gas and/or efficient escape of hydrogen at small orbital 
distances is therefore expected to lead to increased atmospheric mean molecular weights and a wide 
variety of interesting atmospheric compositions. 

The sensitivity-to-metallicity calculations in Fig.~\ref{fig:nepequil} and in most of the investigations 
described in the previous paragraph have been performed assuming that the relative abundances of all 
elements other than H, He, (and sometimes Ne) remain in solar proportion.  Of course, that is not likely going to be 
true for exoplanetary atmospheres, given the different formation scenarios and evolutionary process at work. 
The sensitivity of the chemical equilibrium composition to changes in the C/O ratio at near-solar 
metallicities is discussed in section \ref{subsec:metallicity}.  \citet{moses13gj436b} examine the more 
general case of the sensitivity of the equilibrium atmospheric composition to both the C/O ratio and 
metallicity (i.e., to the relative abundances of H, C, and O) as a function of 
temperature, and \citet{hu14} perform a similar general analysis considering disequilibrium processes like 
photochemistry (see also \ref{subsec:disequilibrium_nep}).  Here is where the diversity of potential 
heavy-element-rich super-Earths and exo-Neptunes really stands out. The dominant equilibrium atmospheric 
constituent on intermediate-sized planets will typically be H$_2$ at low-enough metallicities (e.g., less 
than several hundred times solar) but can become H$_2$O at moderately high metallicities and subsolar C/O 
ratios, can become CO$_2$ at solar-like and subsolar C/O ratios and high metallicities, can potentially 
become CO at high metallicities and C/O ratios near unity (depending on graphite stability), and can even 
become O$_2$ at very low C/O ratios and high metallicities.  At high C/O ratios and low metallicities, 
CH$_4$ is an important atmospheric component at low temperatures, while HCN and C$_2$H$_2$ become more 
important carbon phases at high temperatures.  \citet{hu14} suggest that hydrocarbons like C$_2$H$_2$ 
and C$_2$H$_4$ can even become dominant atmospheric constituents at bulk C/O ratios greater than $\sim$2, 
but they have ignored graphite formation in their calculations.  At high C/O ratios, the graphite stability 
field for equilibrium conditions is greatly expanded \citep[see][]{moses13gj436b}, and for a large range of 
temperature and metallicity conditions, graphite will condense out and sequester a significant fraction of 
the carbon when bulk C/O ratios are greater than $\sim$1 (or even C/O $<$ 1 for high metallicities), leaving 
the remaining gas much less carbon-rich.  Therefore, high-metallicity atmospheres can achieve an unusual 
state where the dominant gas is CO$_2$, even for bulk C/O ratios greatly exceeding unity \citep{moses13gj436b}. It is also possible that atmospheric O may be lost preferentially by photodissociation of H$_2$O and subsequent hydrodynamic escape of O, thereby changing the  C/O ratio \citep{chassefiere1996,luger2015b}.  

Between the potential for hot, silicate- and metal-rich outgassed atmospheres and the variety of volatile 
compositions available from inefficient accretion of nebular gas, the super-Earths and exo-Neptune population 
can be expected to have a rich diversity of atmospheric compositions from thermochemical equilibrium 
considerations, and disequilibrium processes (below) simply augment this possible diversity. 

\subsection{Disquilibrium Chemistry in Exo-Neptune and Super-Earth Atmospheres}
\label{subsec:disequilibrium_nep}

Both photochemistry and transport-induced quenching can affect the atmospheric composition of 
intermediate-size planets, just as on giant planets (see section \ref{subsec:quenching} \& 
\ref{subsec:photochemistry}).  The first photochemical models specifically applied to intermediate-sized
planets were those of \citet{2011ApJ...738...32L} for the exo-Neptune GJ 436b and \citet{miller-ricci12} 
for the super-Earth GJ 1214b, although the generic, higher-metallicity, hot-Jupiter models studied by 
\citet{2009ApJ...701L..20Z,2009arXiv0911.0728Z} should also have relevance to some exoplanets in the 
intermediate-size range.  Thermochemistry, photochemistry, and transport-induced quenching are 
considered in the \citet{miller-ricci12} and \citet{2011ApJ...738...32L} models, and the planets are 
assumed to have H$_2$-rich atmospheres with metallicities up to 30--50$\times$ solar.  Given that
both GJ 1214b and GJ 436b are expected to be relatively cool transiting planets, the results from 
both models are qualitatively similar.  Methane is expected to be the dominant equilibrium carbon 
species for both planets up to 30--50$\times$ solar metallicity for the other conditions 
considered.  At the quench point where interconversion between CH$_4$ and CO shuts down, 
methane is the main carbon component.  However, transport-induced quenching causes CO to be more 
abundant than it otherwise would have been in equilibrium.  This situation represents the opposite of the 
case for hotter planets, where quenching in the CO-dominated regime causes CO to be the major carbon 
component, with methane then being a less-abundant, but still important, quenched component (see 
section \ref{subsec:quenching}).  Photolysis of methane at high altitudes leads to the production of 
C$_2$H$_x$ hydrocarbons and, because of interactions with water photolysis products, the photochemical 
production of CO and CO$_2$.   Coupled methane-ammonia photochemistry causes the production of HCN.  
However, methane is not removed from the photospheric region of either planet due to photochemistry 
(as was suggested as a possibility by \citealt{madhusudhan2011}), which is problematic, given that 
cloudless H$_2$-rich models with equilibrium methane abundances do not reproduce transit and eclipse 
observations of these planets \citep[e.g.,][]{stevenson2010,2011ApJ...731L..40D,2011ApJ...743...92B,2012ApJ...747...35B,kreidberg2014a}.

\begin{figure*}[hbt]
\includegraphics[width=\textwidth]{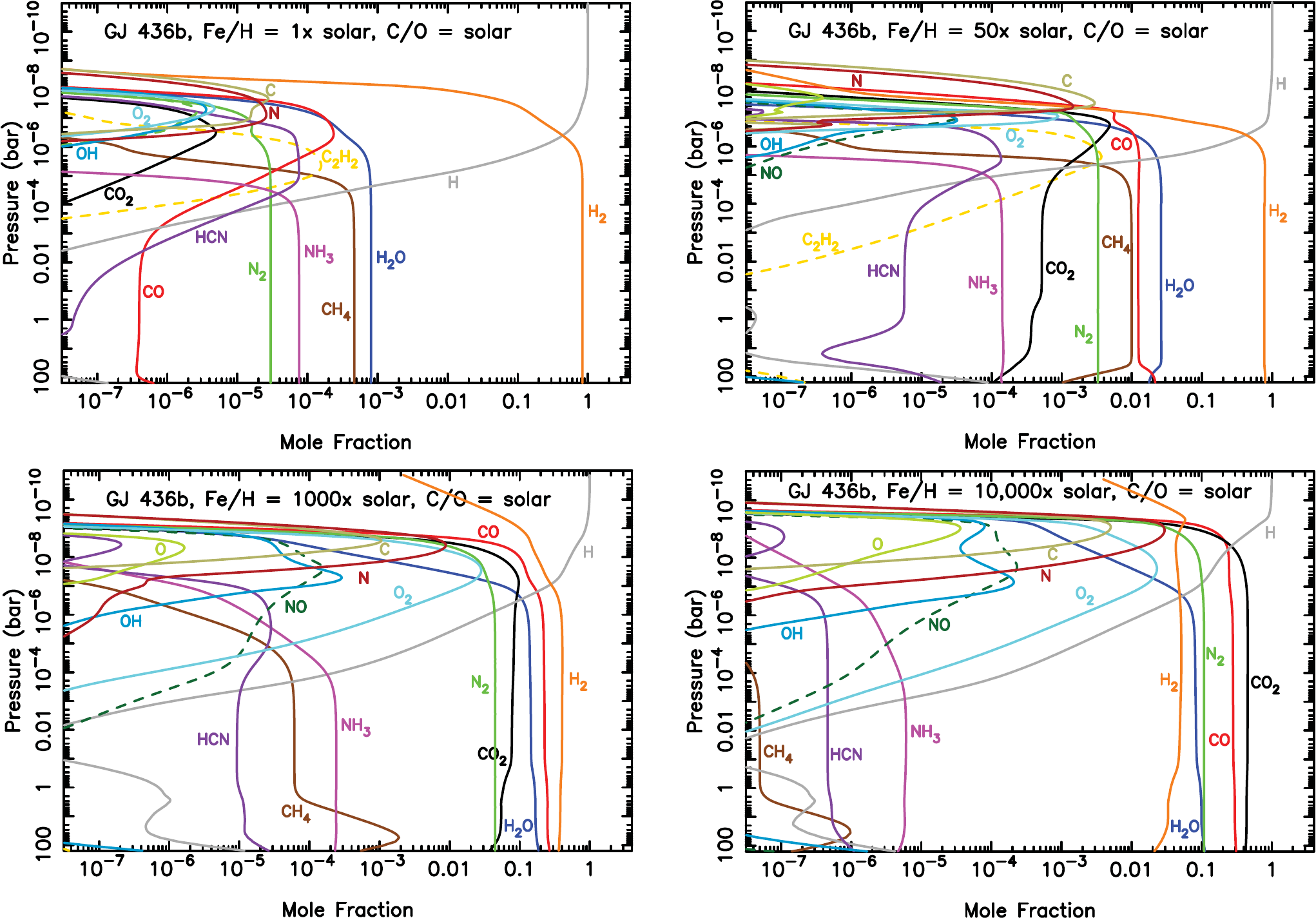}
\caption{Vertical mixing-ratio profiles for several major atmospheric constituents in disequilibrium
thermo/photochemical models of GJ 436b, for assumed metallicities of 1$\times$ solar (top left), 50$\times$
solar (top right), 1000$\times$ solar (bottom left), and 10,000$\times$ solar (bottom right).  Note the
major change in the abundance of CO$_2$, CO, and CH$_4$ with the increase in metallicity (figure from
\citet{moses13gj436b}).
}
\label{fig:pchemgj436b}
\end{figure*}

This model-data mismatch led \citet{moses13gj436b} to suggest that intermediate-sized planets 
could have much higher metallicities than previously considered, as long as their overall H content 
remains consistent with constraints supplied by the planet's bulk density.  As discussed above, higher 
metallicities lead to hotter photospheres at lower pressures and shift the atmosphere toward the CO and 
CO$_2$ stability fields and away from the CH$_4$ stability field.  Figure \ref{fig:pchemgj436b} shows how the 
atmospheric composition and photochemistry change for GJ 436b as the metallicity is assumed to increase 
from 1$\times$ solar to 10,000$\times$ solar.  The atmosphere transitions from being hydrogen-dominated, 
with abundant hydrogen-saturated components like H$_2$O, CH$_4$, and NH$_3$ and photochemically produced 
hydrocarbons and nitriles at low metallicities, to becoming CO$_2$-, CO-, H$_2$O-, and N$_2$-rich, with 
more oxydised photochemical products like O$_2$ and NO at high metallicities.   Condensed graphite is a 
likely cloud component in the very-high-metallicity scenarios.  \citet{moses13gj436b} find that GJ 436b 
models with metallicities in the 230--2000$\times$ solar range provide the best overall consistency with 
both the planet's inferred interior structure \citep[see][and references therein]{nettelmann10} and its 
apparent CO-rich, CH$_4$-poor dayside atmosphere and relatively flat transmission spectrum 
\citep[e.g.,][]{stevenson2010,lanotte2014,knutson2014a}.

\citet{2013A&A...551A.131V} also consider the thermo/photochemistry of a presumed 100$\times$ solar 
metallicity atmosphere for GJ 436b, focusing in particular on how their new measurements of the 
CO$_2$ ultraviolet absorption cross sections at high temperatures (up to 800 K) affect the model 
abundances, and how stellar type affects the results.  \citet{2013A&A...551A.131V} find that their 
results for the predicted abundances of NH$_3$, CO$_2$, and CO are especially affected by the 
temperature-sensitive CO$_2$ cross sections, with models that are irradiated by hotter stars (with 
their corresponding higher near-ultraviolet fluxes) exhibiting the biggest change in abundance due to 
use of high-temperature CO$_2$ cross sections.

Thermo/photochemical models for GJ 436b are also presented by \citet{2014ApJ...781...68A}
and \citet{miguel15}.  \citet{2014ApJ...781...68A} explore the influence of tidal heating and metallicity 
on the thermal structure and resulting disequilibrium chemistry of planet, calculating 
the thermal structure self-consistently.  Both greater tidal heating at depth and higher metallicities 
help shift the thermal structure into the CO dominated regime, again helping favor higher CO/CH$_4$ 
ratios and a greater CO$_2$ abundance.  \citet{miguel15} examine the sensitivity of the atmospheric 
composition of GJ 436b to the flux in the stellar Lyman alpha line for both low-metallicity and 
high-metallicity models.  They find that constituent abundances in the upper stratosphere at pressures 
less than $\sim$0.1 mbar are affected by Lyman alpha, with stronger fluxes favoring the destruction of 
molecules and the resulting dominance of atomic species.  Note, however, that when one considers the 
formation and presence of an extended hot thermosphere \citep{2007P&SS...55.1426G,2013Icar..226.1678K}, 
which is the likely consequence of strong X-ray and EUV radiation received by close-in extrasolar planets, 
--- including GJ 436b \citep[see][]{sanzforcada11,koskinen14esc} --- much or all of the Lyman alpha flux itself 
could be absorbed within the thermosphere \citep{lavvas11}, never reaching the stratosphere.  On the other 
hand, other strong X-ray and EUV lines could have a similar effect to that described in \citet{miguel15}, 
ultimately moving the base of the atomic-dominated hot thermosphere to deeper pressures for stronger EUV 
and X-ray fluxes.  

\begin{figure*}
\centering
\includegraphics[width=0.9\textwidth]{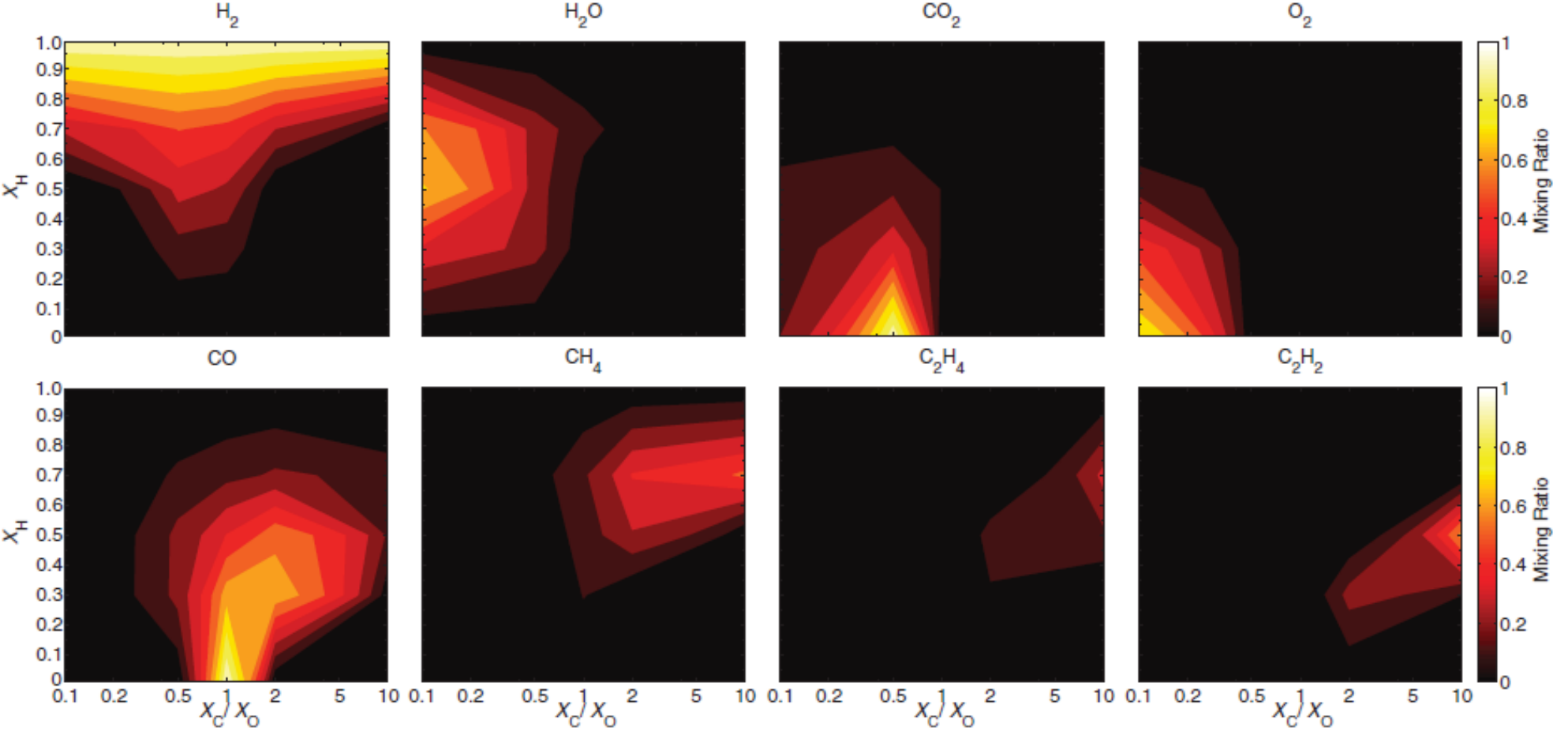}
\caption{Mixing ratios for important atmospheric constituents vertically averaged over the pressure 
range 1--100 mbar from a disequilibrium thermo/photochemistry model of the super-Earth GJ 1214b, as a 
function of the bulk atmospheric H mole fraction ($X_H$, where smaller values correspond to higher 
metallicities) and the C/O ratio ($X_C/X_O$).  Figure from \citet{hu14}.} 
\label{fig:hu14fig5}
\end{figure*}

\citet{hu14} have examined the sensitivity of the disequilibrium atmospheric composition of several 
intermediate-sized exoplanets (e.g., GJ 1214b, HD 97658b, and 55 Cnc e) to the general variation of 
H, O, and C elemental abundances, using a thermo/photochemistry model that self-consistently calculates 
temperatures.  Figure~\ref{fig:hu14fig5} shows their results for GJ 1214b. These results are similar 
to the equilibrium results shown in Fig.~\ref{fig:nepequil} in that the atmospheric composition can be 
highly diverse, depending on the relative abundances of the key volatile elements H, C, and O.  At 
low metallicities, H$_2$ dominates, but GJ 1214b's atmosphere could also be dominated by H$_2$O, CO$_2$, 
CO, or O$_2$, depending on the overall H mole fraction and C/O ratio.  \citet{hu14} suggest that 
the atmosphere could be dominated by hydrocarbons such as C$_2$H$_2$ and C$_2$H$_4$ for high C/O ratios 
and moderately high metallicities; however, graphite condensation has been neglected in their calculations.  
Unless graphite formation is somehow kinetically inhibited \citep[see][]{moses13gj436b}, condensation 
of graphite will tie up much of the excess carbon, causing the remaining vapor to have lower C/O ratios, 
disfavoring hydrocarbons as the dominant constituents.  \citet{hu14} also show how the results for 
intermediate-sized planets can change with stellar flux (and thus temperature), and they suggest 
classification schemes for super-Earths with thick atmospheres based on the dominant constituents 
that appear for different H mole fractions and C/O ratios.  In an earlier study, \citet{hu13} examine 
the fate of sulfur species from photochemistry in atmospheres of different dominant compositions 
relevant to super-Earths.  

\citet{2014A&A...562A..51V} have investigated the disequilibrium chemistry of the warm exo-Neptune 
GJ 3470b, using a grid of disequilibrium models with various metallicities, thermal profiles, incident 
ultraviolet fluxes, and eddy diffusion coefficients (specifying the strength of atmospheric mixing) to 
determine the possible sensitivity of the atmospheric composition to these parameters.  They find that, 
like GJ 1214b and GJ 436b, methane is favored over CO as the dominant atmospheric constituent, except 
when both metallicities (up to 100$\times$ solar) and temperatures are high.

Conspicuously absent from the current literature is a study of how ion chemistry affects the gas-phase 
composition and possible formation of hazes on super-Earths and exo-Neptunes.  Given that ion chemistry 
initiates the formation of high-molecular-weight organics in the N$_2$- and CH$_4$-rich upper atmosphere 
of Titan \citep{waite07}, it seems likely that ion chemistry would be interesting on intermediate-sized 
exoplanets (especially the cooler ones).

\subsection{Clouds and Hazes in Exo-Neptune and Super-Earth Atmospheres}
\label{subsec:clouds_nep}

Basic concepts of cloud and haze formation in extrasolar-planet atmospheres are discussed in section 
\ref{subsec:clouds}.  The same physics and chemistry that was described for giant planets is relevant to 
intermediate-sized planets, but the different starting ingredients can lead to different possible 
aerosol compositions.  \citet{morley2013} provide a good discussion of the possible equilibrium condensates 
along the standard cloud condensation sequence for H$_2$-rich super-Earths and exo-Neptunes.  In order 
from the hottest to the coldest dominant condensates, H$_2$-rich atmospheres could typically contain 
equilibrium clouds of Al-Ca-Ti oxides and silicates, Fe metal, Mg silicates, Cr metal, MnS, Na$_2$S, 
ZnS, KCl, NH$_4$H$_2$PO$_4$, H$_2$O, NH$_4$SH, NH$_3$, and CH$_4$.  For a solar-metallicity atmosphere, 
the total available mass for some of these clouds (e.g., ZnS, KCl) is pretty sparse, suggesting that 
they would not be very optically thick in the vertical; however, \citet{morley2013} demonstrate that 
this conclusion changes as the metallicity is increased.  They find that ZnS and KCl clouds could obscure 
the transit spectra of a metal-rich GJ 1214b if the particles are lofted to sufficiently high altitudes 
and have sedimentation times that are sufficiently long (e.g., due to small particle sizes).  

Other possibilities for aerosols on moderately H$_2$-rich planets include high-molecular-weight organics or 
``soots'' that form through photochemical processes 
\citep[e.g.,][]{2009arXiv0911.0728Z,2011ApJ...737...15M,miller-ricci12,morley2013}, or graphite that is 
stable in equilibrium for metal-rich atmospheres or ones with high C/O ratios \citep{moses13gj436b}.  
Graphite can also be stable for a range of other compositions relevant to outgassed atmospheres of 
intermediate-sized planets \citep{schaefer09,schaefer10}, as can various other equilibrium condensates, 
such as alkali salts \citep{schaefer12}.  Formation of sulfuric acid (H$_2$SO$_4$) clouds is likely through 
photochemical processes under a wide variety of conditions for high-metallicity or CO$_2$-rich atmospheres 
that have sufficient SO$_2$ \citep[e.g.,][]{schaefer11,schaefer12,hu13}, and elemental sulfur aerosols (e.g. S8) can be 
photochemically produced \citep[e.g.,][]{hu13}.  However, the 
detailed formation pathways of photochemical hazes within the diverse atmospheres of intermediate-size 
planets has received relatively little attention and is still poorly understood.

\section{Observational Inferences of Chemical Compositions}	            
\label{sec:inferences}

The last decade has witnessed substantial progress in observational inferences of chemical species in the atmospheres of giant exoplanets through a variety of methods. The planets for which such detections have been made are mostly hot giant planets, either in close-in orbits (i.e. `hot Jupiters') or directly-imaged young objects at large orbital separations. The methods employed include differential photometry and spectroscopy of transiting hot Jupiters, high-resolution doppler spectroscopy of transiting and non-transiting hot Jupiters, and high-resolution spectroscopy of directly-imaged planets. Whereas atomic species have been detected across the ultraviolet and visible, molecular species have been detected primarily in the near infrared. These detections were made thanks to pioneering observations using {\it Spitzer}, {\it HST}, and ground-based facilities. In what follows, we review the developments in each of these various areas. 

\subsection{Chemical Detections via Transit Spectroscopy}
Atmospheric observations have been reported for over 50 transiting exoplanets to date, mostly as broadband/narrowband photometry or low resolution spectra. But, given the limited spectral resolution of the observations of most planets, molecular compositions have been inferred for only a few exoplanets to date. The inferred molecules typically include the most abundant and spectroscopically dominant molecules expected in hot atmospheres e.g. H$_2$O, CO, CH$_4$, and CO$_2$. On the other hand, inferences have also been made of several atomic species and the presence of clouds/hazes in some atmospheres. 

\subsubsection{Detections of Atomic Species} 

In high-temperature gas giant atmospheres (T $\gtrsim$1000 K) in chemical equilibrium, alkali metals sodium (Na) and potassium (K) are expected to exist predominantly in atomic form thereby causing significant absorption in the visible via strong resonance lines at 589 nm and 770 nm, respectively \citep{seager2000,brown2001}. The strong line cores and wide pressure broadened wings are observable in optical transmission spectra of hot Jupiters. Thus the first Na detection was made using a visible band transmission spectrum of the hot Jupiter HD~209458b obtained using the HST STIS spectrograph  \citep{charbonneau2002}. However, while unambiguous, the spectrum revealed lower Na absorption than predicted by equilibrium models, indicating either lower atomic abundances than assumed or non-equilibrium processes such as rainout of condensed species and photoionization of Na/K \citep{fortney2003,barman2007}. 

The Na resonance doublet was also detected from ground in the optical transmission spectrum of the hot Jupiter HD~189733b \citep{redfield2008,wyttenbach2015}. Additionally, HST STIS observations of the transmission spectrum of HD~189733 also detected the Na absorption line core \citep{huitson2012}. However, the ensemble of observations spanning UV to visible showed that the Na line is significantly weaker than that observed for HD~209458b, with a clear lack of broad line wings. Besides the weak Na I line, the spectrum was found to be largely featureless with a blue-ward slope which was consistent with the presence of strong scattering due to a thick haze of condensate grains \citep{vidal-madjar2011,pont2013}. In recent years, Na I has been inferred in visible transmission spectra of a few  other hot Jupiters, e.g. XO-2b \citep{sing2012}, WASP-17b \citep{wood2011,zhou2012}, HAT-P-1b \citep{nikolov2014}, and WASP-12b \citep{burton2015}. 
 
Visible transmission spectra of hot Jupiters have also revealed other atomic species. Similar to Na, several studies have also detected the K resonance double at 770 nm using transmission spectra of hot Jupiters from both space and ground-based instruments, e.g., XO-2b \citep{sing2011a}, and HD~80606 \citep{colon2012}, HAT-P-1b  \citep{wilson2015}, WASP-31b \citep{sing2015}. Other atomic species inferred in the visible include H $\alpha$ at 656 nm \citep{jensen2012}, Ca and Sc \citep{astudillo2013}. 

Several atomic species have also been detected in the exospheres of hot Jupiters using UV transmission spectroscopy. The first such observations in Ly $\alpha$ revealed an extended envelope of escaping H in the hot Jupiter HD~209458b \citep{vidal-madjar2003}. Subsequent observations have revealed a rich population of atomic species in several hot Jupiter exospheres, e.g. H \citep{etangs2010,bourrier2013}, O and C \citep{vidal-madjar2003}, Si \citep{linsky2010,schlawin2010}, Mg \citep{fossati2010,vidal-madjar2013}. Most recently, \citet{ehrenreich2015} reported a detection of escaping atomic hydrogen in the exosphere of the hot Neptune GJ~436b. 
 
\subsubsection{Early Molecular Inferences} 

Early inferences of molecules in exoplanetary atmospheres were based on few channels of photometry or low-resolution spectra obtained using then available instruments on {\it Spitzer} and {\it HST}. For example, some early studies used 2-3 near-infrared photometric observations in transmission to infer the presence of H$_2$O at the day-night terminator regions of HD~209458b \citep{barman2007} and HD~189733b \citet{tinetti2007}, but cf. \citet{beaulieu2008} and \citet{desert2009}. Early attempts were also made to detect molecules in a handful of hot Jupiters using infrared spectroscopy with {\it HST} and {\it Spitzer}. For example, several studies used near-infrared spectra of hot Jupiters in transmission and/or emission obtained with the {\it HST} NICMOS instrument (1.8 -- 2.3 $\mu$m) to report detections of H$_2$O, CH$_4$, CO, and/or CO$_2$ in the hot Jupiters HD~189733b \citep{swain2008a,swain2009a,madhusudhan2009}, HD~209458b \citep{swain2009b}, and XO-1b \citep{tinetti2010}. However, the uncertainties on the NICMOS observations have since been extensively debated in the literature leading to different molecular detection significances claimed by different teams, ranging from confident detections to no detections at the nominal 3-$\sigma$ significance \citep{gibson2011,gibson2012,waldmann2013,swain2014}. 

Similar molecular inferences were also made using observations over a longer spectral baseline using {\it Spitzer} photometry and spectroscopy. \citet{grillmair2008} used {\it Spitzer IRS} spectroscopy to infer the presence of H$_2$O in the dayside atmosphere of HD~189733b using chemical equilibrium models (also see \citet{todorov2014}), whereas \citep{madhusudhan2009} were able to place only an upper-limit on the possible abundances of H$_2$O using the same dataset. Using six-channel photometric and/or spectroscopic observations from {\it Spitzer} and/or {\it HST}, \citep{madhusudhan2009} also inferred the presence of H$_2$O, CH$_4$, CO, and/or CO$_2$ in the hot Jupiters HD~189733b and HD~209458b. However, several of the {\it Spitzer} photometric observations have also since been revised drastically \citep{knutson2012,diamond-lowe2014}, thanks to our improved understanding of the instrumental systematics and, hence, the observational uncertainties \citep[but cf.][]{hansen2014}. Therefore, early inferences based on such observations are not currently substantiated. 

The above early inferences, were followed by more molecular inferences from multi-band photometry for smaller and/or more distant planets. \citet{stevenson2010} and \citet{madhusudhan2011} used six-channel {\it Spitzer} photometry in thermal emission to infer the presence of CO and/or CO$_2$ and the absence of CH$_4$ in the dayside atmosphere of the hot Neptune GJ~436b. Given the relatively lower equilibrium temperature of the planet, the presence of CO and absence of CH$_4$ was suggested to be indicative of strong chemical disequilibrium and high metallicity in the atmosphere. While these data, and hence the molecular inferences, were originally contested by \citet{beaulieu2011}, recent comprehensive and independent analyses by \citet{lanotte2014} support the original inferences of a methane poor and CO/CO$_2$ rich atmosphere in GJ~436b. 

Another major observational advancement that followed was the possibility of detecting thermal emission from hot Jupiters in the near-infrared from ground \citep[e.g.][]{sing2009a,croll2010}. The advantage of these measurements was that their spectral range, between $\sim$0.9$\mu$m and 2.3$\mu$m, complemented that typically available from {\it Spitzer} photometry (3.6 - 8 $\mu$m) to provide a long spectral baseline to retrieve chemical species from the combined data. In the first such instance, observations of thermal emission in seven photometric bandpasses (between $\sim$ 1.2-8 $\mu$m) were used to infer the presence of significant CO and CH$_4$ and lack of H$_2$O, and hence a C/O ratio $\geq$1, in the hot Jupiter WASP-12b \citep{madhusudhan2011a}. Again, the {\it Spitzer} photometric observations were a subject of intense debate initially \citep{crossfield2012,cowan2012}. However, subsequent multi-epoch {\it Spitzer} observations in the same bandpasses have reinstated the original data and the conclusions \citep{stevenson2014a} for the dayside atmosphere of WASP-12b.

\begin{figure}[]
\begin{minipage}{0.5\textwidth}
\centering
\includegraphics[width = 0.95\textwidth]{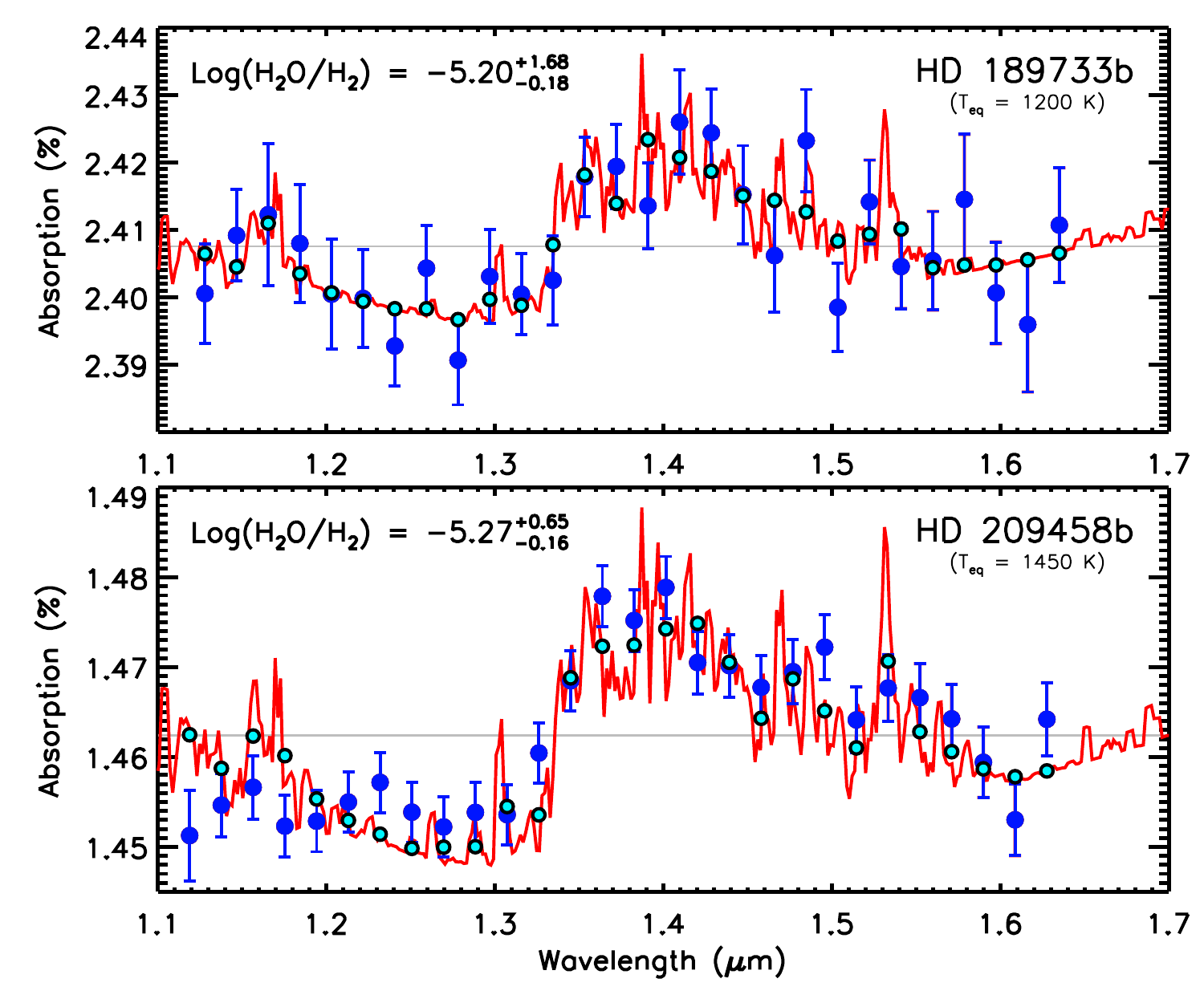}
\caption{Transmission spectra of two hot Jupiters observed with HST WFC3 \citep[adapted from][]{madhusudhan2014a}. The vertical axis shows absorption (transit depth). The blue circles show the data: HD~209458b from \citet{deming2013} and HD~189733b from \citet{mccullough2014}. The red curves shows the best-fit model spectra, and the cyan circles show the models binned to the same resolution as the data. The peaks around 1.4 $\mu$m show H$_2$O absorption.} 
\label{fig:spectra}
\end{minipage}
\begin{minipage}{0.5\textwidth}
\centering
\includegraphics[width = 0.9\textwidth]{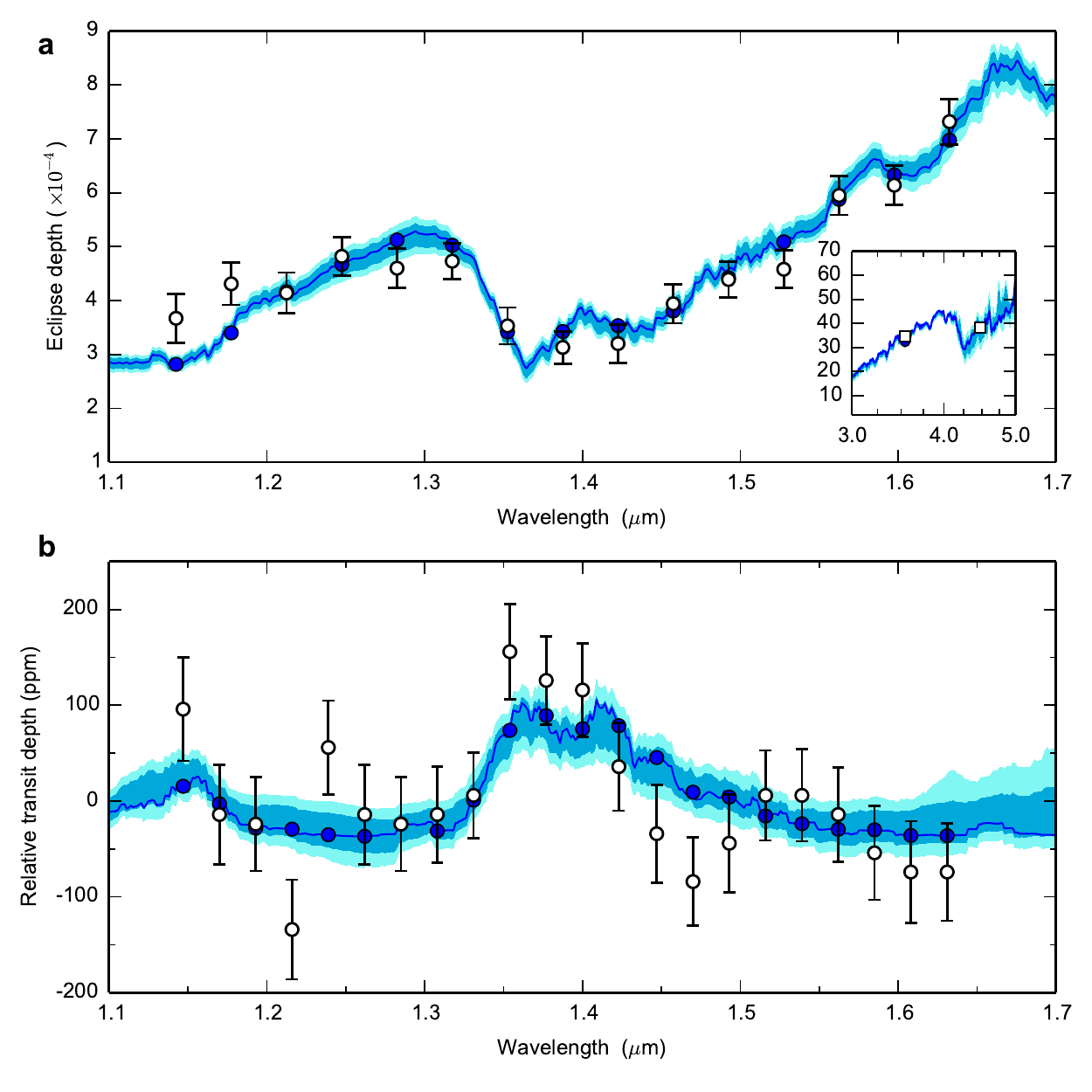}
\caption{Spectra of the hot Jupiter WASP-43b observed with HST WFC3 \citep[from]{kreidberg2014b}. Top: Eclipse spectrum. Bottom: Transmission spectrum. In both cases, the black circles with error bars show the data and solid curves show best-fit model spectra. The filled dark blue circles show the best-fit model binned to the same resolution as the data.} 
\label{fig:h2o}
\end{minipage}
\end{figure}

\subsubsection{Molecular detections with HST WFC3} 

In recent years, near-infrared {\it HST} spectroscopy has led to detections of H$_2$O in the atmospheres of several transiting hot Jupiters. This has been made possible by the {\it HST} Wide Field Camera 3 (WFC3) spectrograph \citep{mccullough2012} which operates in the near-infrared range 1.1-1.8 $\mu$m, where water vapor has a strong absorption band. Since H$_2$O is one of the most abundant molecules expected in giant exoplanetary atmospheres, the HST WFC3 instrument provides a unique opportunity to constrain H$_2$O abundances in such atmospheres. \citet{deming2013} reported the first WFC3 detection of H$_2$O in the transiting hot Jupiter HD~209458b using transmission spectroscopy, which probed the atmosphere at the day-night terminator region of the planet. The observations led to a clear 10-$\sigma$ detection of H$_2$O in a single transit thanks to the high brightness of the host star (V = 7.6).

Several subsequent studies reported single-event transmission spectra of transiting hot Jupiters orbiting less brighter stars with varied levels of success in H$_2$O detections. Some of the notable H$_2$O detections, albeit less significant than that for HD~209458b, were reported for hot Jupiters WASP-17b \citep{mandell2013}, WASP-19b \citep{huitson2013}, HAT-P-1b \citep{wakeford2013}, WASP-12b \citep{stevenson2014a,stevenson2014b}, and HD~189733b \citep{mccullough2014}, and a hot Neptune HAT-P-11b \citep{fraine2014}, whereas similar observations of several other planets resulted in featureless spectra within the observed uncertainties \citep[e.g.][]{swain2013,mandell2013,ranjan2014}. These efforts demonstrate that single transits are inadequate to make high-confidence H$_2$O  detections for most hot Jupiters currently known with the exception of those transiting the brightest stars such as HD~209458b. Consequently, more recent observations using multiple visits to reduce the observational uncertainties have led to good quality transmission spectra with clear H$_2$O detections for hot Jupiters orbiting even moderate brightness stars \citep{kreidberg2014a,kreidberg2015}. 

Several studies have also used {\it HST} WFC3 to observe thermal emission spectra of hot Jupiters at secondary eclipse. In addition to the presence of molecular abundances, thermal emission spectra also provide constraints on the disk-averaged temperature profile of the dayside atmosphere. However, with the exception of a few cases WFC3 observations to date have generally revealed thermal spectra with subdued molecular features, if any. The first such spectrum was observed for the very hot Jupiter WASP-12b (T$_{\rm eq}$$\sim$2500 K) which revealed a featureless thermal spectrum consistent with a blackbody spectrum \citep{swain2013, madhusudhan2012, stevenson2014a}. Considered on its own, the spectrum is consistent with an isothermal temperature structure in the atmosphere. On the other hand, considering together with existing photometric observations in the range 0.9-8 $\mu$m the observations require an atmosphere with a temperature profile decreasing outward and significantly depleted in H$_2$O as expected for a carbon-rich atmosphere, C/O $\geq$ 1, \citep{madhusudhan2012, stevenson2014a}. Subsequent studies have observed WFC3 thermal emission spectra for several other hot Jupiters \citep{wilkins2014,ranjan2014,stevenson2014a,crouzet2014,haynes2015}. 

Robust detections of molecular features in thermal emission spectra using HST WFC3 have been reported for the dayside atmospheres of only two hot Jupiters to date. \citet{stevenson2014c} and \citet{kreidberg2014b} reported unambiguous detection of H$_2$O in the dayside atmosphere of WASP-43b. Most recently, \citet{haynes2015} reported a detection of H$_2$O and the first detection of TiO in the dayside atmosphere of the extremely irradiated hot Jupiter WASP-33b. 

\subsubsection{Abundance Estimations using Retrieval Methods} 

Going beyond detecting the presence of molecules, the exoplanetary spectra discussed above have also been used to derive the molecular abundances which in turn can be used to constrain atmospheric processes, elemental abundances, bulk compositions, and formation conditions. Molecular abundances are derived from exoplanetary spectra using detailed atmospheric retrieval methods which lead to joint statistical constraints on the chemical composition and temperature profile of an exoplanetary atmosphere given an observed spectrum; see e.g. \citet{madhusudhan2014a} for a detailed review of atmospheric retrieval techniques for exoplanets. Atmospheric retrieval methods for exoplanets typically comprise of a 1-D atmospheric model coupled with an optimization algorithm to estimate the free parameters of the model given the data. The molecular abundances and the pressure-temperature ($P$-$T$) profile are free parameters in the model; typically there are over ten free parameters depending on the number of molecules included and the adopted parametrization of the temperature profile. For the optimization algorithm, a number of methods have been tried over time ranging from grid-search in the early days \citep{madhusudhan2009} to Bayesian approaches such as the Markov Chain Monte Carlo (MCMC) method in subsequent years \citep{madhusudhan2011a,line2012,benneke2013,waldmann2015} and gradient descent methods \citep{lee2012}. 

Chemical abundances have been retrieved for several giant exoplanets to date. The majority of these planets are transiting hot Jupiters for which high-precision transmission and/or emission spectra have been obtained using the HST WFC3 spectrograph and, in some cases, Spitzer and/or ground-based photometry. Initial statistical constraints on atmospheric abundances \citep[e.g.][]{madhusudhan2009,lee2012,line2012} in the pre-WFC3 era were based on photometric data and/or low-resolution spectra, the data and uncertainties of which have since been revised as discussed above. On the other hand, while robust HST WFC3 spectra have been observed for over a dozen planets now, most of them have uncertainties large enough to be consistent with featureless spectra \citep{mandell2013,ranjan2014}. Therefore, in what follows we review only a selection of observed transiting exoplanets for which non-flat spectra have been observed at $\geq$3-$\sigma$ significance and for which  abundance constraints have been reported. This includes five transiting hot Jupiters (HD~209458b, HD~189733b, WASP-12b, WASP-33b, WASP-43b), and a hot Neptune (HAT-P-11b). 

Currently, the molecule with the best abundance estimates available is H$_2$O, thanks to high-precision spectra with HST WFC3. As discussed above, the HST WFC3 G141 grism with a spectral range of 1.1-1.8 $\mu$m covers a strong H$_2$O absorption band near 1.4 $\mu$m thereby allowing H$_2$O measurements from exoplanetary spectra. H$_2$O has been detected in all the six giant exoplanets listed above, albeit with different abundances. One important feature in almost all the transmission spectra with robust H$_2$O detections to date is that the amplitude of the H$_2$O absorption feature is typically smaller than model predictions for a clear (i.e. free of clouds/hazes) solar-abundance atmosphere. This was first noted for the hot Jupiter with the most-precise  transmission spectrum, HD~209458b \citep{deming2013}. For such spectra, assuming a clear atmosphere results in an H$_2$O abundance at the day-night terminator that is markedly sub-solar: 20-100$\times$ sub-solar for HD~209458b and 3-200$\times$ sub-solar for HD~189733b \citep{madhusudhan2014b}. However, the derived low H$_2$O abundances are degenerate with the possibility of clouds or hazes at the terminator in these atmospheres \citep{sing2013,benneke2015}, i.e. the same spectra can be explained with solar or super-solar abundances if high altitude clouds or hazes are allowed in the atmosphere. Transmission spectra of several other hot Jupiters were also found to be consistent with solar abundances when clouds/hazes are invoked or when the uncertainties are large enough to preclude high-precision abundance estimates \citep{huitson2013,line2013a,barstow2014,benneke2015}. Therefore, the possibility of clouds present a fundamental challenge in deriving chemical abundances from transmission spectra, which may be mitigated if optical transmission spectra are available (discussed in the following section) and/or where abundance estimates are available for the dayside atmospheres using thermal spectra. 

Constraints on molecular abundances have also been reported in the dayside atmospheres of hot Jupiters using thermal emission spectra. An early example in this regard is the hot Jupiter WASP-12b, for which broadband photometric observations originally suggested a 3-$\sigma$ upper-limit on the H$_2$O abundance that was $\sim$100$\times$ sub-solar, and implied a C/O ratio of $\geq$ 1 \citep{madhusudhan2011a}. While the veracity of the Spitzer photometric observations of WASP-12b were a subject of substantial debate in the intervening years \citep{cowan2012,crossfield2012}, latest multi-epoch Spitzer observations together with high-precision HST WFC3 spectra are consistent with the lack of significant H$_2$O in the dayside atmosphere of WASP-12b \citep{madhusudhan2012,stevenson2014b}. Most recently, H$_2$O was detected at the day-night terminator region of the planetary atmosphere using a HST WFC3 transmission spectrum, however the abundance of H$_2$O detected was subject to model assumptions as discussed above \citep{kreidberg2015}. Similar to the other datasets discussed above, a cloud-free atmosphere required a sub-solar H$_2$O abundance whereas the inclusion of clouds/hazes and/or the imposition of chemical equilibrium allowed the data to be consistent with solar abundance H$_2$O. 

In another example, joint constraints on the H$_2$O abundance at both the terminator as well as the dayside atmosphere were obtained using transmission and emission spectra for the hot Jupiter WASP-43b resulting in an H$_2$O abundance of 0.3-3$\times$ solar, i.e. both sub-solar and marginally super-solar abundances are allowed by the data \citep{kreidberg2014b}. Finally, abundance estimates for the dayside atmosphere of WASP-33b, the most irradiated hot Jupiter observed with HST WFC3 to date, also revealed a marginally sub-solar ($\sim$0.5$\times$) H$_2$O abundance \citep{haynes2015}. HST WFC3 thermal spectra for a few other hot Jupiters have also revealed relatively low-amplitude or non-existent H$_2$O features, hinting at  the possibility of low H$_2$O abundances \citep[e.g.][]{wilkins2014,crouzet2014}. While in principle clouds or hazes could also cause featureless (blackbody-like) spectra in thermal emission, the observed brightness temperatures of the spectra are quite high, indicative of the lower atmospheres, implying that they were unlikely to have been impeded by clouds/hazes in the atmosphere. 

The observed molecular abundances can be used to constrain the O/H, C/H, and C/O ratio of the atmosphere \citep[e.g.][]{madhusudhan2011a}, which in turn are important to constrain exoplanetary formation conditions, as discussed in section~6. However, estimating C/O ratios is non-trivial. Reliably estimating the C/O ratio for an exoplanetary atmosphere requires that the observations are able to constrain the molecular abundances of all the dominant O and C bearing molecules in a given atmosphere. As discussed in section 3, for hot giant exoplanets, the dominant molecules are H$_2$O, CO, CH$_4$, and to a lesser extent CO$_2$, C$_2$H$_2$, and HCN, depending on the C/O ratio. Simultaneously constraining the abundances of all these molecules requires a long spectral baseline with observable bandpasses where these molecules have strong spectral features. This is, in principle, possible with the combination of HST WFC3 and the Spitzer IRAC Channels 3.6-8 $\mu$m \citep{madhusudhan2012}. Indeed, such constraints were placed for the few planets where such data are available in thermal emission allowing constraints on the C/O ratios of their dayside atmospheres. While a carbon-rich dayside atmosphere (C/O $\geq$ 1) has been reported for the hot Jupiter WASP-12b \citep{madhusudhan2012,stevenson2014b}, oxygen-rich dayside atmospheres consistent with a solar C/O of 0.5 were reported for the hot Jupiters WASP-43b \citep{kreidberg2014b} and WASP-33b \citep{haynes2015}. 

On the other hand, determining the C/O ratio at the day-night terminator from transmission spectra is challenging and prone to misinterpretation. One of the main reason is that typically only WFC3 spectra are available in transmission, implying that stringent constraints are possible only on the H$_2$O abundance. In principle, the abundances of CH$_4$ and HCN may also be constrained in the WFC3 spectral range, however the most dominant O and C bearing molecule in hot Jupiters is CO, which is unconstrained by WFC3 observations. Consequently, with WFC3 data alone the C/H and hence the C/O ratio cannot be reliably constrained. For example, if a sub-solar H$_2$O abundance is observed in a hot Jupiter (assuming a cloud-free atmosphere), the low H$_2$O can be caused either by a low overall metallicity (i.e. low O/H) with a solar-like C/O ratio or by a high metallicity but a high C/O ratio ($\sim$1), depending on the temperature. Therefore, studies which attempt to derive C/O ratios from WFC3 transmission spectra alone do so by enforcing constraints of equilibrium chemistry or temperature profiles in radiative equilibrium or both \citep{benneke2015,kreidberg2015}, neither of which is strictly justified at the day-night terminator regions of strongly irradiated atmospheres with strong atmospheric circulation. On the other hand, the presence of clouds only makes the inference more degenerate. Therefore, multi-wavelength observations besides WFC3 spectra are essential for reliable constraints on C/O ratios from transmission spectra. 


\begin{figure}[]
\centering
\includegraphics[width = \textwidth]{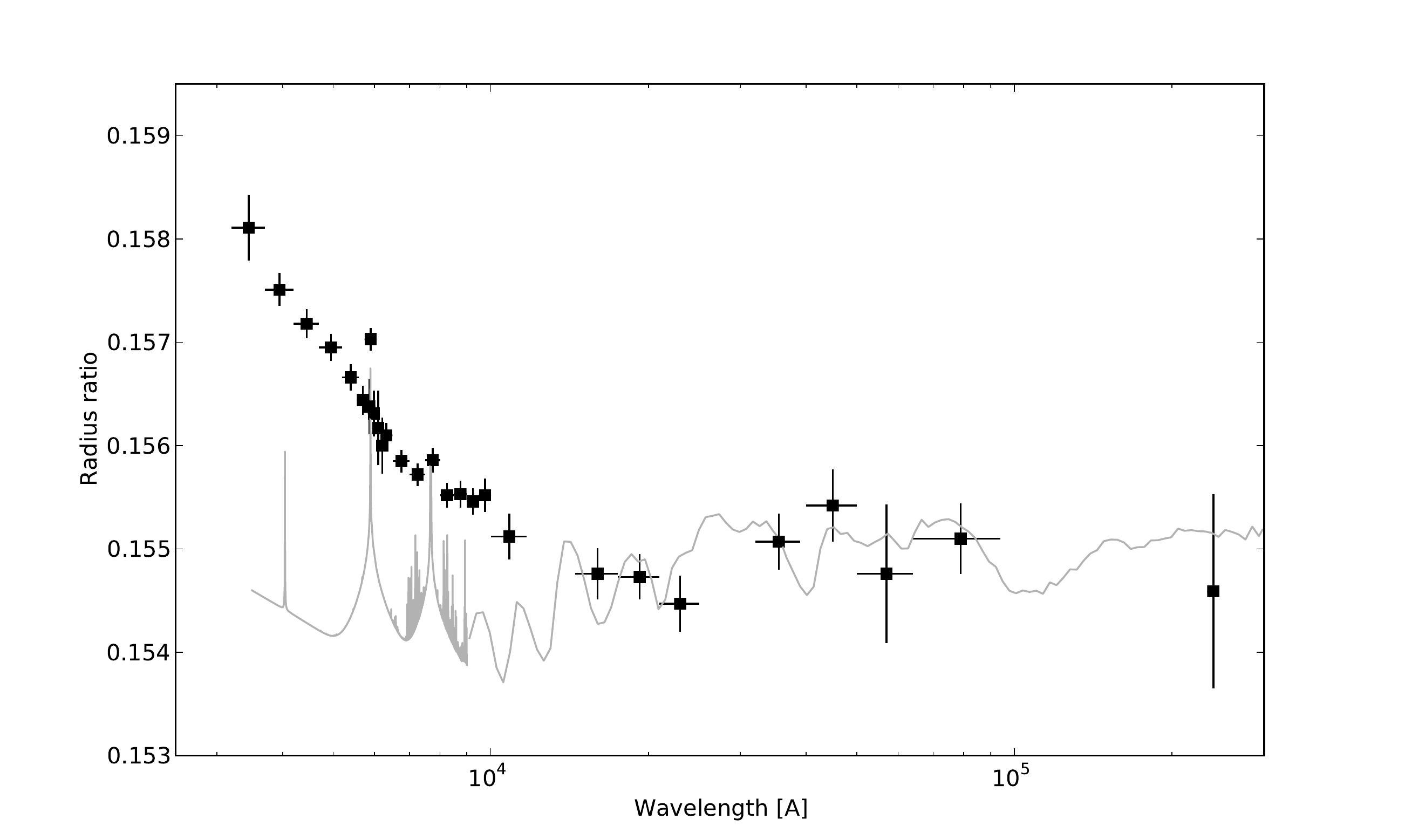}
\caption{Transmission spectrum of the hot Jupiter HD~189733b from UV to near-infrared \citep[from][]{pont2013}. The black circles with error bars show the data available at that time, observed using multiple instruments on {\it HST} and {\it Spitzer}. The grey curve shows a model spectrum for a dust-free atmosphere. The steep blue-ward rise in the spectrum was indicative of strong scattering due to haze/dust in the atmosphere which was not consistent with the expected model spectrum for a haze-free atmosphere.}
\label{fig:pont}
\end{figure} 

\subsubsection{Degeneracy with Clouds and Hazes} 

The presence of high-temperature condensates as hazes and clouds in hot exoplanetary atmospheres may contribute significantly to the optical depth in transmission spectra \citep{seager2000,fortney2005}. Therefore, determinations of chemical compositions from transmission spectra can be strongly degenerate with the presence of clouds or hazes in the atmospheres. In transmission spectra, the amplitude of a spectral feature is directly proportional to the height of the atmospheric annulus through which the star light traverses. The presence of an opaque cloud deck at a given height in the atmosphere means that only the layers above the cloud deck contribute to the transmission spectrum. Thus for a cloud deck located high enough in the atmosphere the amplitude of a spectral feature can be significantly diminished thereby confounding estimates of the corresponding molecular abundances. 

High-precision observations of several transiting exoplanets have revealed such spectra with diminished features. At the extreme end are recent HST WFC3 observations of remarkably flat spectra, despite extremely high precisions, for super-Earths GJ~1214b \citep{kreidberg2014a} and HD~97658b \citep{knutson2014b}, exo-Neptune GJ~436b \citep{knutson2014a}, and exo-Uranus GJ~3470b \citep{ehrenreich2014}. The featureless transmission spectra in these low-mass, low-temperature (T below $\sim$ 800 K), planets have been attributed to the presence of thick high-altitude clouds with cloud-top pressures below a mbar. On the other hand, even for hot Jupiters (T$\sim$1200-3000 K) which show unambiguous H$_2$O features the amplitude of the spectral features are significantly diminished, corresponding to only $\sim$1-2 scale heights \citep{deming2013,kreidberg2015}, compared to expectations of 5-8 scale heights \citep{madhusudhan2015}. These diminished spectral features may be attributed to the presence of clouds/hazes in these hot Jupiters, but as discussed above the same features may also be attributed to significantly lower H$_2$O abundances than previously expected \citep{madhusudhan2014a,benneke2015}. 

Scattering due to haze particles have also been inferred from visible transmission spectra of hot Jupiters which show an inverse power-law dependence on wavelength with parameters different from that due to pure gaseous Rayleigh scattering. The archetype for such inferences is the hot Jupiter HD~189733b \citep{etangs2008,pont2008,sing2011b} where observations with HST revealed a power law slope in the UV and optical indicating the presence of hazes. Similar inferences have been made for several other hot Jupiters in recent years \citep{sing2013,sing2015,nikolov2015}. A potential degeneracy in such inferences arises from the fact that the presence of star spots may also cause similar optical spectra with inverse power law slopes \citep[e.g.][]{mccullough2014}. However, in addition to the power-law slopes observations of partially muted alkali (Na and/or K) absorption lines in the optical have also indicated the presence of cloud decks in some atmospheres  \citep{sing2013,sing2015}. 

\subsubsection{Albedos and Optical Phase Curves} 
\label{sec:albedos}

The presence of clouds/hazes have also been inferred using reflection spectra of hot Jupiters obtained at occultation. Early observations presented little evidence for clouds/hazes and suggested that hot Jupiters are dark, potentially due to strong absorption in the optical due to alkali metals Na and K. Such studies reported rather low geometric albedos for several hot Jupiters in the optical, e.g. 0.038 $\pm$ 0.045 for HD~209458b \citep{rowe2008} and  0.0136 $\pm$ 0.0027 for TrES-2 \citep{kipping2011}. However, recent new observations are providing evidence for high geometric albedos and evidence for clouds/hazes in dayside spectra of hot Jupiters. \citet{demory2011,demory2013} reported a high visible geometric albedo of 0.35 $\pm$ 0.02 for the hot Jupiter Kepler-7b observed in the Kepler photometric bandpass ($\sim$0.4-1.0 $\mu$m). Furthermore, based on the asymmetry in the visible phase curve they also suggested the presence of an inhomogeneous cloud cover in the atmosphere of Kepler-7b. In another study, \citet{evans2013} reported a low-resolution albedo spectrum of the hot Jupiter HD~189733b in the 0.29-0.57 $\mu$m range using HST which nominally showed the albedo increasing blue-ward and inferred the presence of reflective clouds in the atmosphere; also see \citet{barstow2014}. Most recently, \citet{martins2015} reported a high-resolution spectroscopic detection of reflected light for the non-transiting planet 51 Peg b which suggests a high geometric albedo ($\sim$0.5) of the planet. Several other recent studies have used photometric observations to understand the trends and/or report constraints on albedos and/or optical phase curves for a sizeable sample of exoplanets \citep[e.g.][]{heng2013, esteves2015, hu2015a, schwartz2015}. 

While it is apparent from the above observations that clouds/hazes likely exist in at least a subset of irradiated atmospheres, the compositions of cloud forming species in currently unconstrained. However, as discussed in sections 3-4, theoretical studies over the past decade show that a wide range of high-temperature refractory condensates can form in such atmospheres in the 500-2000 K range. Such condensates range from NaCl, KCl, ZnS for $T$ below 1000 K to silicates, Fe, and Al$_2$O$_3$ at $\sim$1000-2000 K temperatures \citep[e.g.][]{sudarsky2003,marley2013,morley2013}. 

\subsection{Molecular Detections via High-resolution Doppler Spectroscopy} 

Robust detection of molecules in exoplanetary atmospheres have also been made using high-resolution 
doppler spectroscopy in the near infrared. This technique involves the detection of molecular lines in the planetary 
spectrum that are shifted in wavelength due to the radial velocity of the planet \citep{brogi2012,birkby2013}. For close-in hot Jupiters, the orbital velocities are $\sim$km~s$^{-1}$ whereas the stellar orbital velocities are significantly lower 
(e.g. below $\sim$100 m~s$^{-1}$). Thus the spectrally shifted molecular lines of the planetary atmosphere are easily identifiable 
compared to those in the stellar spectrum as well as those in the telluric spectrum which is static. A template planetary 
spectrum including the sought after molecule is cross-correlated with the observed spectrum to detect the doppler shift 
in the molecular lines with phase thereby revealing the presence of the molecule. Critical to this method, however, is the 
high resolution of the observed spectrum so that individual molecular lines can be resolved. Consequently, most of the success in this area has been achieved using the CRIRES instrument on the Very Large Telescope (VLT) with a spectral resolving power of 100,000 in the near-infrared \citep{snellen2010}. Moreover, as the planetary signal diminishes with the increased resolution the method has been successfully applied only to planets orbiting the brightest stars. 

This method has been used to robustly detect CO and H$_2$O in several hot Jupiter atmospheres. The detections of these 
particular molecules are favored by the fact that they are expected to be the most dominant O and C bearing molecules in hot Jupiter atmospheres, especially for $T > 1300$~K, and also have detectable absorption lines in the range $\sim$2-3 $\mu$m where these atmospheres are most conducive to observe from ground. This wavelength range offers the optimal conditions because in the near-infrared the planet-star flux ratio increases with wavelength, however the background noise also increases with wavelength; therefore the K-band (around 2.1 $\mu$m) typically offers an optimal choice for ground based transit/eclipse spectroscopy. \citep{snellen2010} reported the first detection of CO using this 
technique in the hot Jupiter HD~209458b in transit which in turn also led to a constraint on the day-night wind velocity in the planetary atmosphere. Subsequent observations have led to the detection of CO and H$_2$O in the dayside atmospheres of several transiting and non-transiting planets: $\tau$Boo \citep{brogi2012,rodler2012,lockwood2014}, 51 Peg b \citep{brogi2013}, HD~189733b \citep{dekok2013,birkby2013,rodler2013}, and HD~179949b \citep{brogi2014}. 


\subsection{Directly Imaged Gas Giants} 

Direct imaging offers another avenue to characterize atmospheric compositions of exoplanets and works preferentially 
for a complementary region in planetary parameter space. Exoplanets detected to date via direct imaging 
are all young gas giants at large orbital separations. Their young ages (below $\sim$100 Myr) implies high effective 
temperatures ($\sim$1000 K) and, hence, high planet-star flux 
contrasts in the near-infrared. On the other hand, their large orbital separations (beyond $\sim$10 AU), which are set by 
the inner working angles of the instruments used, help in minimizing the stellar glare. Together these factors make young 
gas giants particularly favorable to near-infrared spectroscopy via direct imaging. 

The advantage of direct imaging is that the detection of a planet simultaneously results in observation of its atmospheric 
thermal emission spectrum, i.e. atmospheric characterization 
can be pursued simultaneously with detection. This is in contrast to transiting exoplanets where the detections are generally made 
using photometric transit surveys and RV confirmations, while atmospheric spectra are obtained for optimal targets using follow-up 
observations with specialized instruments. On the other hand, the challenge with directly imaged planets is that generally only the 
atmospheric spectrum is available with little information about any other planet property; the mass, radius, and hence gravity, temperature, 
age, are all unknowns in modeling the planets. Therefore, typically, planetary evolution models are required along with atmospheric 
models to robustly constrain the atmospheric and physical parameters of the planets. Moreover, spectra are available at only one point 
of the orbital phase which means thermal phase curves are not observable precluding constraints on atmospheric properties with longitude. 
Nevertheless, the possibility of high-resolution absolute near-infrared spectroscopy for directly-imaged planets means that chemical signatures 
can be robustly detected in emission spectra using ground-based instruments. 

Chemical detections have been reported for a few directly-imaged exoplanets in recent years. Given the expected temperature range 
of young giant planets, the dominant molecular species expected are H$_2$O, CO, and CH$_4$. All these molecules have been detected in one or 
more directly-imaged planets. Most of the detections have been reported for planets in the well characterized and nearby HR 8799 system. 
H$_2$O, CO, and CH$_4$ have been unambiguously detected in the planet HR 8799b using high-resolution spectroscopy in the H and K 
band \citep{barman2011a,barman2015}, though the CH$_4$ absorption is weaker than expected from chemical equilibrium \citep{bowler2010,barman2015}. 
H$_2$O and CO have also been detected in high-resolution spectra of HR 8799c \citep{konopacky2013}. The presence of CH$_4$ and other candidate molecules (e.g. NH$_3$, C$_2$H$_2$, CO$_2$, and/or HCN) have also been suggested using lower resolution spectra for the four planets in the HR~8799 system \citep{oppenheimer2013}. 
\citet{janson2013} reported an H-band detection of CH$_4$ in GJ 504b. More recently, CO was also detected unambiguously in $\beta$ Pic b \citep{snellen2014} which also provided the first measurement of the spin period of an exoplanet. 

In addition to molecular detections, spectra of directly imaged planets have also been use to place constraints on their elemental abundance ratios just as have been pursued for transiting hot Jupiters. However, determining elemental abundance ratios for directly imaged planets is challenging since most of the planetary properties (e.g. mass, radius, gravity, temperature) are not known a priori and hence need to be fit for, often requiring evolutionary models, in addition to the usual complexities introduced by the presence of clouds. Nevertheless, several studies have attempted to fit static models to spectra of directly-imaged planets to report nominal constraints on the elemental abundances. Using a model grid to fit spectroscopic observations of HR~8799 c, \citet{konopacky2013} suggested a C/O ratio of $0.65^{+0.10}_{-0.05}$ which is slightly higher than a solar value of 0.54. \citet{lee2013} performed an atmospheric retrieval analysis for HR~8799b to suggest a high metallicity, with a mean molecular weight of 3.8 compared to 2.4 for solar abundance, and a C/O ratio $\sim$ 1 in its atmosphere. More recently, \citet{todorov2015} performed an atmospheric retrieval analysis to constrain the molecular abundances in the sub-stellar companion $\kappa$~And~b and reported an H$_2$O abundance of $\sim$$10^{-4}$, nearly consistent with a solar abundance composition and those of several transiting hot Jupiters. 

\begin{figure}[]
\centering
\includegraphics[width = 0.8\textwidth]{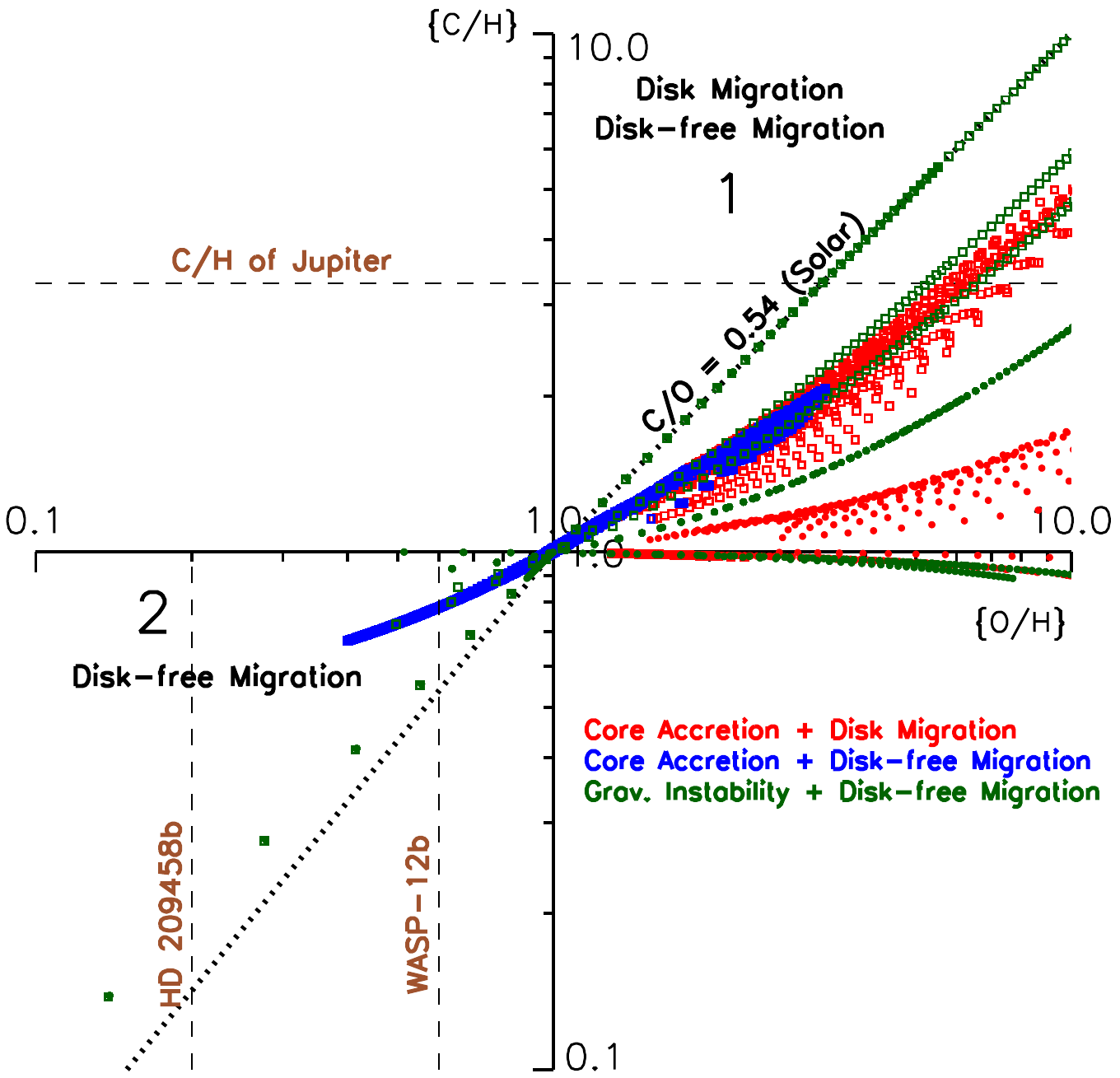}
\caption{Predicted elemental O and C abundances in hot Jupiters from semi-analytic chemical models of hot Jupiter formation and migration \citep[from][]{madhusudhan2014c}. The axes show O/H and C/H abundances relative to solar values, and the colored symbols represent models with different formation-migration histories. Each dot represents a different model realization and shows the final O/H and C/H of the planet resulting from that model. The final elemental abundances in a planet depend critically on the amount of gas and solids accreted by the planet as well the region from which that material is accreted relative to the H$_2$O, CO$_2$, and CO snow lines, giving rise to the distinct loci of models in the O-C plane as shown. The results suggest that different migration mechanisms can cause different patterns of elemental enhancements in hot Jupiters.}
\label{fig:formn}
\end{figure} 

\section{Implications for Formation Conditions of Exoplanets} 
\label{sec:formation}
Atmospheric elemental abundances of solar-system giant planets have led to important constraints on the origins of the solar system. In Jupiter's atmosphere, the abundances of C, N,S, Ar, Kr, and Xe, relative to H have been measured to be 2-3$\times$ solar values \citep{owen1999,atreya2005} which suggest substantial accretion of solids during its formation, and have been used as evidence for its metal-rich interior and its formation by core accretion. However, the abundance of O is not known for Jupiter. Given the low temperatures ($\lesssim$ 125 K) in Jupiter's observable troposphere, H$_2$O is condensed down to deeper layers of the atmosphere at pressures greater than $\sim$10 bar. As the Galileo probe descended into Jupiter's atmosphere, its last measurement showed an H$_2$O abundance, and hence an O abundance, of ~0.3$\times$ solar, which is 10$\times$ lower than the other elements \citep{atreya2005}. However, this measurement of Jupiter's H$_2$O abundance is considered to be a lower-limit since the region the probe entered is thought to be an anomalously dry spot in Jupiter's atmosphere. On the other hand, indirect constraints based on kinetics models explaining the observed  CO abundance suggest an O abundance of 0.3-7.3 $\times$ solar \citep{visscher10co}, which allows for a wide range of C/O ratios. 

Taking the currently available lower-limit on O at its true abundance, implying more carbon than oxygen, would require unusual formation conditions in the early solar system. \citet{lodders2004} studied the hypothetical possibility and suggested that tarry planetesimals may have dominated Jupiter's accretion history, instead of planetesimals dominant in water-ice as expected from compositions of minor bodies in the solar system. Instead, using core accretion models \citet{mousis2012} suggested that considering a solar-composition nebula predicts an O abundance of 3-7$\times$ solar. Therefore, a significantly lower observed O abundance would require substantial depletion of H$_2$O ice in Jupiter's formation region, unless the planetesimals are unusually carbon-rich. Thus, accurately measuring the O abundance is critical to constrain the formation conditions of Jupiter, and of the outer solar system in general. The H$_2$O abundance is similarly unknown for any other giant planet in the solar system. The upcoming Juno mission to Jupiter \citep{matousek2007}, therefore, aims to constrain its O abundance by measuring the H$_2$O abundance in Jupiter's atmosphere. In particular, the H$_2$O abundances play a central role in constraining exoplanetary formation conditions. Since O is cosmically the most abundant heavy element it is expected that H$_2$O is one of the most dominant volatile in interstellar and planet-forming environments \citep{vandishoeck2014}. Thus, the H$_2$O abundances in giant planetary atmospheres are of fundamental importance to constrain the inventory of O that was available in planet's formation environments.

The O/H and C/O ratios are easier to measure for hot giant exoplanets than they are for solar-system giant planets \citep{madhusudhan2012}. The vast majority of extrasolar gas giants known have T $\sim$ 600-3000 K, thus hosting gaseous H$_2$O in their observable atmospheres accessible to spectroscopic observations. Other detectable gases include CH$_4$, CO, CO$_2$, and NH$_3$, depending on the temperature and incident irradiation. Measurements of such molecular abundances allow estimations of elemental abundances ratios involving H, C, O, and N. Such elemental abundances can in turn provide crucial clues regarding exoplanetary atmospheric processes, interior compositions, and formation mechanisms, just as pursued for solar system planets. This fortuitous opportunity makes hot giant exoplanets the perfect laboratories to investigate the origins of giant planets and the diversity of their atmospheres and interiors. Indeed, as discussed above, recent  observations are already leading to unprecedented detections of chemical species in giant exoplanetary atmospheres. Such spectra have already led to clear H$_2$O detections in atmospheres of several transiting hot Jupiters, as discussed above \citep[e.g.][]{deming2013,kreidberg2014b,madhusudhan2014b}. H$_2$O, CO, and/or CH$_4$ have also been detected using high-resolution ground-based spectroscopy of hot Jupiters \citep[e.g.][]{brogi2014}, as well as directly imaged planets \citep[e.g.][]{konopacky2013}.  Upcoming large facilities such as the {\it JWST} and {\it E-ELT} will further revolutionize the field. 

New studies are beginning to investigate the influence of formation and migration histories of giant exoplanets on their observable chemical compositions. 
As discussed in previous sections, observations are suggesting the possibility of carbon-rich (C/O $\geq$ 1) as well as oxygen-rich (C/O $<$ 1) atmospheres in giant exoplanets \citep{madhusudhan2011a,madhusudhan2012} which are in turn motivating new ideas on the formation conditions of planetary systems. One of the key questions in this regard is about how C-rich gas giants can form around O-rich sun-like stars; the solar C/O is 0.5 \citep{asplund2009}. As discussed above, an early investigation into this question was pursued in the context of Jupiter in the solar-system for which, as discussed above, only a lower limit on the O/H is known, which allows for a C/O $>$ 1 in Jupiter. More recently, following the inference of C/O $\geq$ 1 in WASP-12b \citep{madhusudhan2011a}, \citet{oberg2011} suggested that C/O ratios in giant planetary envelopes depend on the formation location of the planets in the disk relative to the snow lines of major C and O bearing volatile species, such as H$_2$O, CO, and CO$_2$, since the C/O ratio of the gas approaches 1 outside the CO and CO$_2$ snow lines. By predominantly accreting such C-rich gas, more so than O-rich planetesimals, gas giants could, in principle, host C-rich atmospheres even when orbiting O-rich stars. On the other hand, it may also be possible that inherent inhomogeneities in the C/O ratios of the disk itself may contribute to higher C/O ratios of the planets relative to the host stars  \citep{kuchner2005,madhusudhan2011b,mousis2012,moses2013a,ali-dib2014}. The compositions of gas and solids accreted also depend on the physicochemical properties of the disk at the given location which are time-dependent \citep{helling2014,marboeuf2014}. Furthermore, some of the volatile elements can also be depleted due to dust formation and settling in the atmosphere \citep{moses2013a,helling2014}. These various scenarios predict different limits on the metallicites and C/O ratios of the giant planets, which high-precision observations of their atmospheres will be able to test in the near future.

Besides the various factors discussed above, the main formation mechanisms also contribute significantly to the planetary composition. Giant planets are thought to form via one of two primary mechanisms, core accretion (CI) versus gravitational instability (GI). In the CI model \citep{pollack1996}, the planetary embryos start out as ~10 Earth-mass cores in the protoplanetary disk that subsequently undergo runaway accretion of a large volume of gas and planetesimals to form a massive gaseous envelope. On the other hand, a GI in a young disk can cause rapid collapse of a large volume of ambient gas and solids to form a giant planet \citep{boss2000}. Both scenarios occur in planet-forming disks, but at different orbital separations. While CA is favored closer to the snow-line (within $\sim$2-10 AU) because cores take too long to form at larger distances and only reach large masses after the disk has dispersed, GI is favored at larger distances ($\gtrsim$10 AU) where the disk can cool sufficiently on orbital timescales to fragment. In this regard, GI may be the favored mechanism for the formation of distant gas giants detected via direct imaging. However, because the different formation mechanisms are favorable at different orbital separations in the disk they may be expected to probe different chemical compositions in the disk. In principle, both CA and GI can cause significant metallicity enhancements or depletions depending on the specific accretion history \citep{helled2010}, but the relative elemental abundances (e.g. C, O, N, etc.) may be different depending on the formation location corresponding to the respective snow lines. 

The existence of hot Jupiters in very close-in orbits (e.g. $\lesssim$ 0.1 AU) presents a key challenge to theories of giant planet formation. Neither GI nor CA discussed above is thought to operate in such a way that allows hot Jupiters to form in situ at their current locations close to the host stars. The disk cannot fragment at those distances \citep[][]{gammie2001,rafikov2005}, and cores with sufficient mass to attract significant envelopes cannot form. Therefore, the existence of hot Jupiters requires some form of "migration'' from their original formation locations to their present orbits. Two competing hypotheses suggest that the planets migrated either through interaction with the protoplanetary disk during their formation  \citep{papaloizou2007}, or by `disk-free' mechanisms such as gravitational interactions with a third body \citep[e.g.]{rasio1996,fabrycky2007}. Measurements of spit-orbit misalignment (or stellar obliquity) for a large sample of hot Jupiters over the past decade \citep[e.g.][]{triaud2010,albrecht2012} have been advocated as possible metrics to distinguish between the two hypotheses for migration. In a simplistic view, migration through the disk was thought to likely align the orbital angular momentum vector of the planet with the stellar spin axis whereas disk-free migration via dynamical encounters could lead to spin-orbit misalignment. A significant number of large spin-orbit misalignments (or stellar obliquity) observed in hot Jupiter systems initially supported disk-free migration mechanisms \citep{winn2010}.  However, recent studies show that spin-orbit misalignments may also be caused by planet migration through disks that are themselves misaligned and due to star-disk interactions \citep{crida2014,lai2014}. Consequently, observed dynamical properties of hot Jupiters have been unable to conclusively constrain their migration pathways. 

Instead, recent studies suggest that chemical abundances of hot Jupiters could provide stronger constraints on their formation and migration pathways. As discussed above, the O/H, C/H, and C/O ratios can change substantially with their formation locations in the protoplanetary disk. \citet{madhusudhan2014c} suggested that atmospheric metallicities of hot Jupiter atmospheres could potentially constrain their migration mechanisms as migration through the disk is more likely to cause metal enrichment due to planetesimal accretion compared to disk-free migration mechanisms. The results lead to three key predictions as evident from Fig.~\ref{fig:formn}. Firstly, planets migrating through the disk always accrete solids efficiently enough to result in solar or super-solar abundances (top-right quadrant in Fig.~\ref{fig:formn}). Secondly, the C and O abundances are enhanced or depleted together, i.e. no cases in the top-left or bottom-right quadrants. Thirdly, planets with sub-stellar O and C abundances could not  result from disk-migration. These findings imply that elemental abundances of hot Jupiters could potentially provide important constraints not only on their local formation environments but also on the migration pathways. Current observations are already providing the first constraints on the O and C abundances which in turn are being used to constrain formation and migration mechanisms \citep{madhusudhan2014c}. These efforts will be further bolstered in the future as the spectral range of JWST instruments will contain strong molecular features of several molecules which will allow precise abundance estimates for a wider range of elements (e.g. O, C, N, P, S, Si, etc). 

\section{Theory: Climates of Habitable Exoplanets}                     
\label{sec:habitability}

\subsection{Habitability of exoplanets}

Existence of life requires many kinds of conditions. One of the critically necessary conditions for a planet to harbor life is permanent existence of liquid water on the planet's surface. To maintain liquid water, the planetary surface temperature has to be between the freezing point of water ($0^\circ$C) and the boiling point ($100^\circ$C), i.e., at 1 bar, 

\begin{equation}
0^{\circ}\mathrm{C} < T_s < 100^{\circ}\mathrm{C}.
\label{eq:71}
\end{equation}

In reality, the upper limit of surface temperature has to be lower than the boiling point of water. Otherwise, the planet would fall into the runaway greenhouse state. We will return to this issue in the next section. To retain a fairly dense atmosphere and large amount of water against escape processes, the planetary mass has to be large enough so that the planet has sufficiently strong gravity. However, the planetary mass cannot be too large. Otherwise, it could have accreted a massive H$_2$-He envelope whose greenhouse effect along with the high pressure of the envelope would warm the surface and prevent water from being liquid, or it could be a Neptune-like ice-giant planet~\citep{Selsis2007, Venturini2015}. Specifically, a habitable planet has to be a solid planet, and its mass should be in an approximate range given by: 

\begin{equation}
0.5M_{\oplus} < M < 10 M_{\oplus}, 
\label{eq:72}
\end{equation}

where, $M_{\oplus}$ denotes Earth mass \citep{Selsis2007}. 

Based on the above two conditions, an Earth Similarity Index (ESI) has been proposed \citep{Schulze2011}  which attempts to quantify the similarity of a given property of an exoplanet to that of the Earth and ranges between 0.0 and 1.0. There are about 30 exoplanets that have ESI greater than 0.5. In other words, there are about 30 potentially habitable exoplanets among nearly 2000 discovered exoplanets. Gl 581c is the first exoplanet that had ever been thought to be habitable~\citep{Udry2007}. However, radiative-convective model simulations indicated that Gl 581c is too hot to maintain liquid water~\citep{Selsis2007, Hu2011}. Especially, \citet{Hu2011} showed that the climate of Gl 581c can readily fall into runaway greenhouse state even if the CO$_2$ level is extremely low and cloud albedo is extremely high. Gl 581d had also been considered a habitable super-Earth~\citep{Udry2007, Selsis2007, von2007}, and it is probably the first exoplanet whose habitability has drawn intensive studies. \citet{Wordsworth2010}, \citet{Hu2011}, and \citet{von2011} all showed that at least 7 bars of CO$_2$ are required to maintain the surface temperature of Gl 581d above the freezing point of water. Kepler 186f, which has a size similar to Earth's, was also suggested to be a habitable exoplanet. However, calculations indicate that the surface temperature of Kepler 186f would be much colder than that of Gl 581d for same CO2 concentration because Kepler 186f receives less stellar radiation than Gl 581d does. So far, none of these potentially habitable exoplanets has been confirmed to be habitable. 

\subsection{Habitable zones}
Condition (\ref{eq:71}) defines a circumstellar zone in which a terrestrial planet can hold permanent liquid water on its surface. Such a zone is the so-called Habitable Zone (HZ). The HZ was first proposed by~\citet{Shapley1953} as the region around a star in which liquid water could exist on a planet's surface. The term `habitable zone' was first introduced by~\citet{Huang1960}, in the context of planetary habitability and extraterrestrial life. The HZ was more precisely defined in later works~\citep{Kasting1993, Kopparapu2013}, based on results from one-dimensional radiative-convective climate models. In these works, various factors that influence the width of the HZ, such as surface albedo, atmospheric compositions, and stellar radiation spectra, were considered. Especially, \citet{Kasting1993} extensively addressed how the inner and outer edges of HZ are constrained by the positive water-vapor feedback and by the saturation limit of the maximum CO$_2$ concentration, respectively. One might expect that the inner edge of HZ is at the distance where planetary surface temperature is equal the boiling point of water ($100^{\circ}$C). However, as shown by \citet{Ingersoll1969}, all liquid water on the planetary surface would completely evaporate into the atmosphere due to the positive water-vapor feedback once the surface temperature reaches about $70^{\circ}$C, and the planet falls into the runaway greenhouse climate state. Then, water is lost throughout photolysis and hydrogen escape. The outer edge is constrained by the limit of CO$_2$ saturation, with which CO$_2$ condensation begins to take place. Thus, further increasing CO$_2$ does not enhance the greenhouse effect. Instead, latent heat release by CO$_2$ condensation even causes weakened greenhouse effect due to the decrease of lapse rate. Using the one-dimensional climate model, \citet{Kasting1993} found that conservative estimates of the inner and outer edges of HZ in our own Solar System is about 0.95 and 1.37 AU, respectively. Thus, the width of the solar HZ is about 0.42 AU. It is wider than the mean spacing between the four terrestrial planets in the Solar System, which is about 0.35 AU. It suggests that, statistically speaking, at least one of the terrestrial planets ought to be in the solar HZ. \citet{Kasting2010} pointed out that the possibility that at least one rocky planet will be in the HZ should be high if such a "statistical result" can be applied to other exoplanetary systems. For darker stars like M dwarfs that have effective temperatures of about 3500 K, the HZ is much closer to the stars, at about 0.1 AU. 

One uncertainty of the HZ width calculated from radiative-convective models is cloud radiative effects that are not included in these one-dimensional models. It was thought that the inner edge could be pushed closer to stars as the negative feedback of cloud reflection of stellar radiation is considered~\citep{Kasting1993, Selsis2007}. Three-dimensional general circulation models (GCMs) include cloud radiative effects because GCMs have self-adjusted dynamical and physical processes of cloud formation. Several recent works have studied the HZ width with GCMs. However, \citet{Leconte2013} demonstrated that the inner edge of the solar HZ calculated from their GCM is nearly the same as that predicted by one-dimensional models. The reason why the inner edge does not move closer to Sun in their GCM is because of two important factors. One is that water-vapor absorption of solar radiation is higher in their model than previously assumed. The other one is that for high surface temperatures and water-rich atmosphere conditions high cirrus clouds increase in fraction faster than low stratus clouds. The former consists of ice particles and have greenhouse effects, and the later mainly reflects solar radiation and cools the surface. On the other hand, \citet{Yang2013} used a GCM to show that deep convective clouds develop around the substellar point for slow-rotating or tidally-locked exoplanets. These clouds largely reflect stellar radiation and cool the surface. As a result, the inner edge of the HZ is pushed much closer to the star, and the HZ width is nearly doubled. 
	
It is worthwhile to point out that the cloud radiative effect on the inner edge of the HZ has large uncertainties. Parameterized clouds in GCMs are the primary source of uncertainty even in simulating modern climates where we have a wealth of direct observations. Thus, it is not clear to what extent cloud parameterizations in Earth GCMs can be applied to studying exoplanetary atmospheres. Using Earth GCMs to study the outer edge of the HZ also has great challenges. There is also lack of parameterizations of CO$_2$ clouds. Radiative effects of CO$_2$ clouds are not well understood~\citep{Forget1997}. 

\subsection{Climates and habitability of tidal-locking exoplanets in the HZ of M dwarfs}

M dwarfs are the most common stars in the Universe~\citep{Rodono1986}. Thus, the probability of finding habitable exoplanets around M dwarfs is much higher than around other types of stars. Moreover, it is much easier to find habitable exoplanets closer to stars than that with greater distances with current observation techniques. Therefore, it is very likely to first discover habitable exoplanets around M dwarfs. Since M dwarfs have much weaker luminosities than Sun-like stars, the HZ around an M dwarf is typically about 0.1 AU, much closer than that for a solar-type star. However, the short distance of the HZ could also cause exoplanets in the HZ around M dwarfs to be uninhabitable, as discussed below. 

First, exoplanets within such short distance to M dwarfs can be exposed to high levels of X-ray and extreme ultraviolet radiation (X-EUV) and strong particle fluxes from stellar winds or coronal mass ejections~\citep{Lammer2011}. Such strong emissions are due to M dwarfs' magnetic activity. It can generate significant atmospheric escapes to space and can even potentially cause atmosphere erosion or strip the whole atmosphere. For solar-type stars, such strong X-EUV emissions may be limited to the first few hundred millions years. However, the extreme irradiation could last several Gyr for M-type stars. Thus, exoplanets in the HZ of M-type stars may receive X-EUV fluxes that are 10-100 times higher than those in the HZ of solar-type stars of the same age~\citep{Selsis2007}. X-EUV flux could cause a CO$_2$ dominant atmosphere with modest amount of free oxygen (O$_2$). Rapid escape of carbon from a CO$_2$ dominant atmosphere could lead to the formation of O$_2$ even without the presence of biological O$_2$ production \citep{Tian2009}. Photochemical models also suggest that O$_3$ and CH$_4$ can also be produced abiotically for different stellar characteristics and levels of volcanic outgassing \citep{domagal2014}. On the other hand, several other compounds, e.g. of sulfur, could act as potential biosignatures \citep{domagal2011}. At the current stage, more observational and quantitative modeling works are needed for understanding the effects of the extreme irradiation on atmospheres of exoplanets in the HZ around M dwarfs \citep[see e.g.][]{segura2010}. 

Second, habitable exoplanets orbiting M-type stars are very likely to be tidally locked to their primaries due to strong forcing of gravitational gradients. Tidal-locking exoplanets receive very uneven stellar heating because their one side permanently faces their stars and the other side remains dark. While the dayside can be warm enough to sustain liquid water, the nightside could be so cold that any gases condense out there. If there exists atmosphere collapse, exoplanets in the HZ of M dwarfs would be uninhabitable. There were works that studied whether atmospheric heat transports could prevent atmosphere from collapse on the nightside using simplified or sophisticate models\citep{Haberle1996, Joshi1997, Joshi2003, Pierrehumbert2011}. They all showed that for a sufficiently dense atmosphere (surface air pressure greater than 0.1 bar) atmospheric heat transports would be able to warm the nightside and maintain nightside surface temperature above the condensation point of CO$_2$ ($-78.5^{\circ}$C at 1 bar of CO$_2$ partial pressure, and much lower at lower partial pressures). Therefore, atmosphere collapse can hardly happen as long as exoplanets are able to attract a fairly dense atmosphere envelope. However, more recent works suggest that the threshold of air pressure to prevent atmosphere collapse requires further investigation~\citep{Heng2012a, Wordsworth2015}. 

\begin{figure}
  \includegraphics[width=0.9\textwidth]{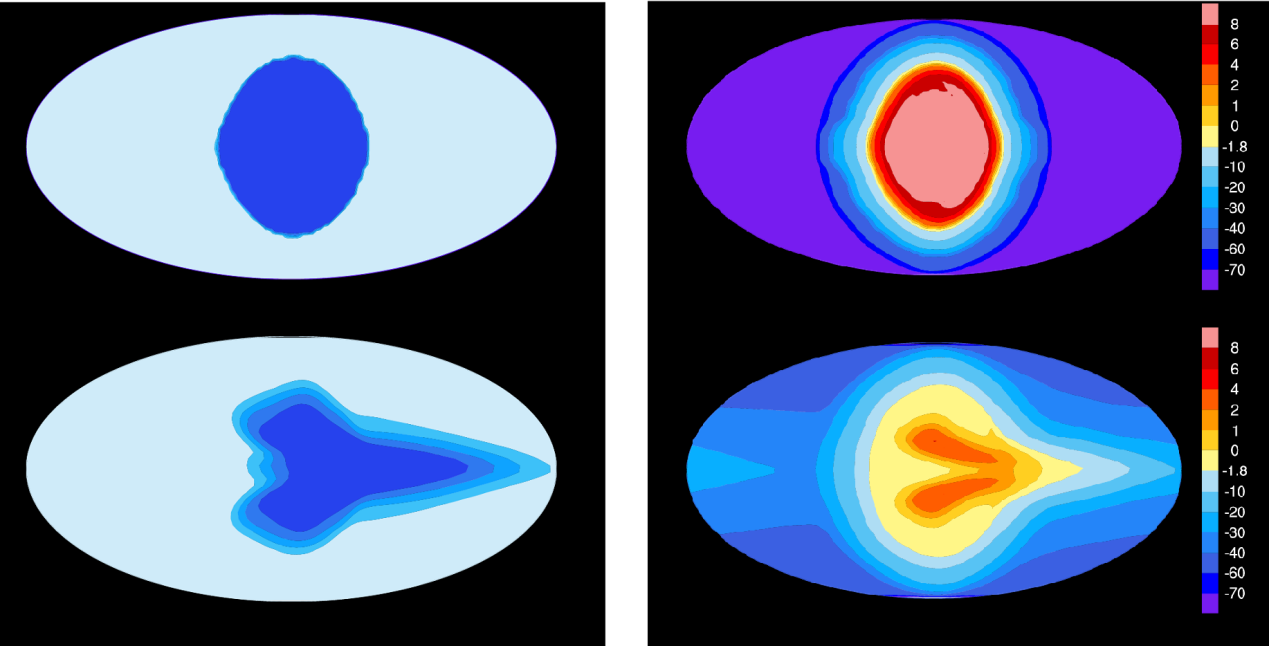}
  \caption{Spatial patterns of open-ocean areas and surface temperatures. Left panels: Sea-ice fraction. Blue colors indicate open-ocean areas, and light-gray indicates ice coverage. Right panels: surface temperatures, color bar unit is $^{\circ}$C. Top panels: with only atmospheric heat transport, and bottom panels: heat transports by both atmosphere and ocean. The substellar point is at the center of each panel. After Hu and Yang (2014).}
  \label{fig:71}       
\end{figure}

The above studies have not taken ocean into account. If an exoplanet has an extensive ocean, its habitability also involves ocean heat transports and the positive sea-ice albedo feedback~\citep{Hu2014}. On the one hand, it is well known that ocean heat transports are equally important in Earth's climate~\citep{Peixoto1992}. In the presence of sea ice, ocean heat transports are likely to be especially important, since it is known from studies of the Snowball Earth phenomenon in Earth-like conditions that ocean heat transports are very effective in holding back the advance of the sea-ice margin~\citep{Pierrehumbert2011, Yang2012c, Yang2012a, Yang2012b}. On the other hand, in the presence of sea ice there exists the possibility that the exoplanet could be locked in a globally glaciated Snowball state due to the positive ice-albedo feedback. Note that the distribution of ice on tidally locked exoplanets is a particularly interesting issue only for M-type stars, since exoplanets orbiting hotter stars in orbits close enough to yield tidal-locking are likely to be too hot to permit ice, and may even be too hot to retain liquid water. Using a fully coupled atmospheric-oceanic GCM (AOGCM), \citet{Hu2014} carried out simulations and demonstrated that ocean heat transports substantially extend the area of open water along the equator, and that the open-ocean area shows a lobster-like spatial pattern (Figure \ref{fig:71}), instead of an eyeball as shown in \citet{Pierrehumbert2011}. Figure \ref{fig:71} also shows that as CO$_2$ concentration increases, the dayside temperature does not increase very much, but the nightside temperature increases a lot, indicating that most heat is transported from the dayside to the nightside by ocean currents. They also showed that ocean heat transports can even lead to complete deglaciation of the nightside as greenhouse gas concentration is sufficiently high or stellar radiation is sufficiently strong. By contrast, the open-ocean area is not expanded very much by atmospheric heat transports alone for the same high level of CO$_2$. They also showed that it is more ready for aqua-planets to fall into runaway greenhouse and runaway freezing at the inner and outer edges, respectively, compared with simulation results without a dynamic ocean. It implies that the HZ would be narrower for aqua-planets or exoplanets with extensive oceans. These results indicate that ocean heat transports play critically important roles in determining the climate state and habitability of exoplanets. 

\begin{figure}
  \includegraphics[width=0.9\textwidth]{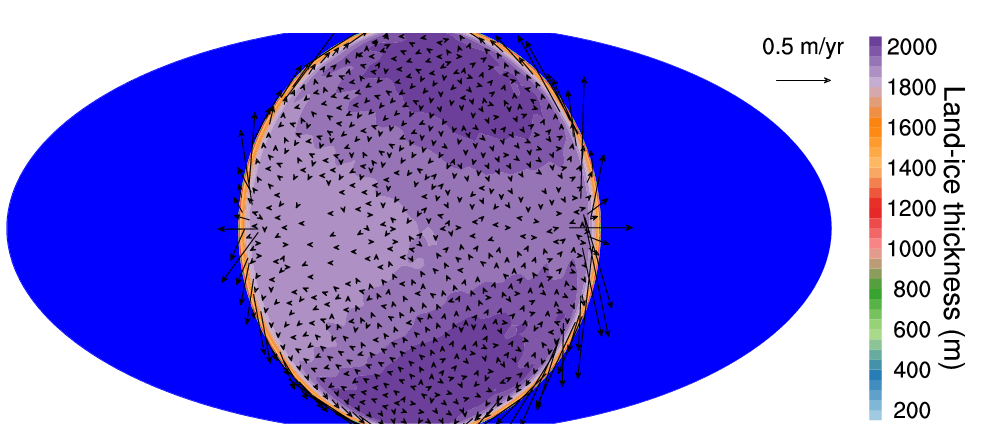}
  \caption{Ice sheets over a supercontinent spanning the entire nightside. Color shading indicates ice-sheet thickness, and arrows indicate ice-sheet flow velocity. Geothermal flux is 0.1 $W m^{-2}$ in the simulation. 
  Adapted from \citet{Yang2014}. 
 }
  \label{fig:72}
\end{figure}

The third problem is whether water could be completely trapped on the nightside of tidal-locking exoplanets in the HZ of M dwarfs~\citep{Menou2013}. Although Figure \ref{fig:71} shows that the lowest surface temperature on the nightside is about $-40^{\circ}$C, well above the condensation point of CO$_2$, it is still far below the freezing point of water. Therefore, there is the possibility of water trap on the nightside. \citet{Hu2014} showed that as long as the ocean is not completely frozen, wind stresses transport sea ice toward the dayside and the ocean carries heat toward the nightside. As a result, sea-ice thickness on the nightside remains thin, less than 10 meters. \citet{Yang2014} studied that a case with a super-continent on the nightside and an ocean on the dayside. In this case, ocean heat transports from the dayside to the nightside are ceased. It is found that ice sheets over the nightide supercontinent can grow 2000 m thick if geothermal heat flux is close to Earth's or smaller (Figure \ref{fig:72}). It suggests that if the dayside ocean is deeper than 2000 m, water will not be completely trapped on the nightside continent. Only exoplanets with a geothermal heat flux lower than Earth's, and much of their surface covered by continents would be susceptible to complete water trapping. 

\subsection{Observable signatures of tidally-locked habitable exoplanets}
The above climate patterns have important implications for future observations. \citet{Showman2009, Showman2011,heng15} showed that an equatorial superrotating jet stream develops in the atmosphere of tidal-locking exoplanets. The hot spot of tidal-locking exoplanets is advected to the downstream of the substellar point by the equatorial superrotating jet stream. Their simulation results are confirmed by observations for hot Jupiters~\citep{Knutson2009}. It was shown that the hot spot is shifted $20^{\circ}$-$30^{\circ}$ east of the substellar point of hot Jupiter HD 189733b. The equatorial superrotating flow is a common feature for tidal-locking exoplanets~\citep{Heng2011, Hu2013}. Thus, the hot-spot shift shall also be observed from the thermal phase curve of terrestrial exoplanets. In fact, the climate pattern in Figure \ref{fig:71} also suggests eastward shift of the hot spot. Especially, as a dynamic ocean is considered, the equatorial ocean current, together with the equatorial atmospheric jet stream, can cause further downstream shift of the hot spot. \citet{cowan2012b} studied how surface thermal inertia of Earth-like exoplanets leads to distinguishable thermal phase curve for observations. They studied two types of climate states. One is the temperate climate which is just like the modern Earth with low- and middle-latitude oceans and polar ice-caps. The other one is the Snowball state. They found that the former has a relatively flat thermal phase curve because its relatively large surface thermal inertia tends to damp out the amplitude of the thermal phase curve. \citet{Wang2014} argued that terrestrial exoplanets around M dwarfs could have different climate states. They use an atmospheric GCM coupled with a slab ocean to show that exoplanets with nonzero eccentricities could have different spin-orbit resonance states different from the synchronous rotation state, and these exoplanets should have a striped-ball climate pattern, with a global belt of open ocean at low and middle latitudes and ice caps over both polar regions. This is in contrast to synchronous rotating habitable exoplanets around M dwarfs that have an eyeball climate pattern a limited region of open water on the dayside and ice on the rest of the planet~\citep{Pierrehumbert2011, Hu2014}. They pointed out that the striped-ball climate state should be common for habitable exoplanets in large eccentric orbits around M dwarfs. They further suggested that these different climate patterns can be observed by future exoplanet detection missions. Overall, the features of climate patterns provide future observations of habitable exoplanets with additional constraints and information. 

\section{Future Prospects} 
\label{sec:future}

Major observational advancements are happening in two key directions: (a) detections of exoplanets around bright stars, and (b) high precision observations of exoplanetary spectra. Firstly, bright exoplanet host stars are important to obtain precise measurements of exoplanetary masses, radii, and atmospheric spectra. Upcoming space missions, CHEOPS \citep{broeg2014} and TESS \citep{ricker2015}, as well as various ground-based surveys (e.g. NGTS, SPECULOOS, MEARTH, etc.) are expected to find thousands of transiting exoplanets, including hundreds of super-Earths, orbiting nearby bright stars within the next five years (also see chapter on future landscape by Fridlund et al. 2016 in present issue). During the same time, major ground-based direct imaging platforms (e.g. SPHERE, GPI, etc.) are also expected to find at least dozens of exoplanets at wide orbital separations orbiting nearby young stars. On a longer run, early 2020s, the recently selected ESA mission, PLATO \citep{rauer2014}, will discover numerous transiting exoplanets in habitable zones of nearby stars. Secondly, major parallel efforts are also being pursued for spectroscopic observations of exoplanetary atmospheres. In this regard, enormous amount of time is currently being dedicated on the Hubble and Spitzer space telescopes, as well as major ground-based telescopes (e.g. VLT, Keck, Gemini, Magellan, CFHT, etc.). In addition, the James Webb Space Telescope ({\it JWST}) scheduled for launch in 2018 will revolutionize exoplanetary spectroscopy. By early 2020s, next generation ground-based telescopes, such as the European-Extremely Large Telescope (E-ELT), will further revolutionize the field. Future spectra with {\it JWST} would be of unprecedented precision and resolution which will enable us to derive precise chemical abundances for transiting exoplanets. In addition, JWST would also be able to observe atmospheres of much cooler low-mass planets that are possible to observe today. \newline\newline

{\it Acknowledgements:} This review was conceived during the 2014 ISSI workshop in Beijing on ``The Disk in Relation to the Formation of Planets and their Protoatmospheres". NM and MA thank the International Space Science Institute (ISSI) and ISSI-Beijing for the hospitality during the workshop and the participants of the workshop for the lively discussions on exoplanetary atmospheres. MA thanks K. Lodders for interesting discussions on atmospheric condensates. M.A. acknowledges funding support from Spanish MINECO through grants CSD2009-00038, AYA2009-07304, and AYA2012-32032 and from the European Research Council (ERC Grant 610256: NANOCOSMOS). JM thanks the NASA Exoplanet Research program NNX15AN82G for support. YH is supported by the National Natural Science Foundation of China 435 (NSFC) under grants 41375072 and 41530423. 

\bibliographystyle{aps-nameyear}      

\bibliography{issi_moses,issi_refs_NM,issi_Hu} 

%
%
%

\end{document}